\documentclass[12pt,a4paper,oneside]{article}
\usepackage{graphicx,amssymb,amsmath,color}
\usepackage[linkcolor={blue},citecolor={red},colorlinks=true]{hyperref}
\usepackage{enumerate}
\usepackage[margin=2cm]{geometry}
\usepackage{slashed}
\usepackage[normalem]{ulem}

\def\a{\alpha}

\def\beq{\begin{equation}}
\def\eeq{\end{equation}}
\def\bea{\begin{eqnarray}}
\def\eea{\end{eqnarray}}

\def\bit{\begin{itemize}}
\def\eit{\end{itemize}}

\def\l{\left}
\def\r{\right}

\def\la{\lambda}
\def\s{\sigma}

\def\baa{\begin{array}}
\def\eaa{\end{array}}

\def\d{\partial}

\def\simgt{\mathrel{\lower2.5pt\vbox{\lineskip=0pt\baselineskip=0pt
           \hbox{$>$}\hbox{$\sim$}}}}
\def\simlt{\mathrel{\lower2.5pt\vbox{\lineskip=0pt\baselineskip=0pt
           \hbox{$<$}\hbox{$\sim$}}}}
\newcommand{\vev}[1]{ \langle {#1} \rangle }

\def\bfc{\begin{figure}\begin{center}}
\def\efc{\end{center}\end{figure}}
\def\nn{\nonumber\\}

\def\e{\epsilon}
\def\f{\phi}
\def\G{\Gamma}
\newcommand{\laq}[1]{\label{eq:#1}}

\newcommand{\AND}{~{\rm and}~}
\newcommand{\EV}{ {\rm ~eV} }

\newcommand{\GEV}{ {\rm ~GeV} }
\newcommand{\TEV}{ {\rm ~TeV} }
\def\O{\mathcal{O}}
\def\({\left(}
\def\){\right)}

\begin{document}

\begin{flushright}
\hspace{3cm} 
SISSA 03/2021/FISI
\end{flushright}
\vspace{.6cm}
\begin{center}

\hspace{-0.4cm}{\Large \bf 
Dark Matter production from relativistic bubble walls}\\[0.5cm]

\vspace{1cm}{Aleksandr Azatov$^{a,b,c,1}$, Miguel Vanvlasselaer$^{a,b,c,2}$ and Wen Yin$^{d,3}$}
\\[7mm]
 {\it \small

$^a$ SISSA International School for Advanced Studies, Via Bonomea 265, 34136, Trieste, Italy\\[0.15cm]
$^b$ INFN - Sezione di Trieste, Via Bonomea 265, 34136, Trieste, Italy\\[0.1cm]
$^c$ IFPU, Institute for Fundamental Physics of the Universe, Via Beirut 2, 34014 Trieste, Italy\\[0.1cm]
$^d$ Department of Physics, The University of Tokyo,   
Bunkyo-ku, Tokyo 113-0033, Japan\\[0.1cm]
 }

\end{center}

\bigskip \bigskip \bigskip

%%%%%%%%%%%%%%%%%%%%%%%%%%%%%%%%%%%%%%%%%%%%%%%%%%%%%%%%%%%%%%%%%%%%%%%%%%
\centerline{\bf Abstract} 
\begin{quote}

In this paper we present a novel mechanism for producing the observed Dark Matter (DM) relic abundance during the First Order Phase Transition (FOPT) in the early universe. We show that the bubble expansion  with ultra-relativistic velocities can lead to the abundance of DM particles with masses much larger than the scale of the transition. We study this non-thermal production mechanism in the context of a generic phase transition and the electroweak phase transition. The application of the mechanism to the Higgs portal DM as well as the signal in the Stochastic Gravitational Background 
are discussed.\\

\end{quote}

\vfill
\noindent\line(1,0){188}
{\scriptsize{ \\ E-mail:
\texttt{$^1$\href{mailto:aleksandr.azatov@NOSPAMsissa.it}{aleksandr.azatov@sissa.it}},
\texttt{$^2$\href{miguel.vanvlasselaer@NOSPAMsissa.it}{miguel.vanvlasselaer@sissa.it}},
\texttt{$^3$\href{yinwen@tuhep.phys.tohoku.ac.jp}{yinwen@tuhep.phys.tohoku.ac.jp}}
}},

\newpage

\newpage
\section{Introduction}

Cosmological observations conspire to suggest the existence of a massive, undetected, dark component permeating the universe\cite{Bertone:2016nfn}, this is the Dark Matter (DM) phenomenon. One of the earliest candidate for this DM, the celebrated WIMP component, demands that the Standard Model (SM) is coupled to the DM, whose stability is guaranteed by a symmetry.  This interaction leads to quick thermalisation between the DM and the SM. In this mechanism, known as \emph{thermal Freeze-Out} (FO), thermal relic density is naturally fixed via the decoupling of the SM-DM sectors, when the rate of interaction can not compete any more with the expansion of the universe\cite{Srednicki:1988ce, GONDOLO1991145, Griest:1990kh}. The requirement that this relic density matches the observed abundance imposes a relation between the DM-SM coupling and the mass of the DM candidate. In this context, the surprising and exciting coincidence that weak coupling and TeV scale DM candidate are consistent with observed DM abundance is known as the \emph{WIMP miracle}.
Moreover, unitarity considerations on the coupling governing the scattering of DM provide an upper bound on the mass of the DM candidate\cite{PhysRevLett.64.615}, the \emph{Griest-Kamionkowski (GK) bound} of $O(100)$ TeV. 
However, today, many WIMP models have been excluded due to the bounds on the DM-nucleon scattering set by the direct detection experiments\cite{Agnese:2015nto,Akerib:2016vxi,Tan:2016zwf,Angloher:2015ewa,Amole:2017dex}.

To diversify the range of possibilities inside the (coupling-mass) parameter space, many alternatives to FO have been proposed, as for example; freeze-in \cite{Hall:2009bx, McDonald:2001vt, Chu:2013jja}, forbidden freeze-in\cite{Darme:2019wpd}, super-heavy particles decay \cite{Chung:1998zb, Allahverdi:2018iod}. Several proposals also take advantage of the possibility of an early First-Order Phase Transitions (FOPT) occurring in the universe, with many different consequences on DM abundance \cite{SCHRAMM198553}. Phase transitions offer a way to fix the final relic abundance via the VEV flip-flop mechanism \cite{Baker:2016xzo,Baker:2018vos, PhysRevD.101.043527}, by modifying the stability of DM candidate \cite{Cohen:2008nb,Bian:2018mkl}, through the injection of entropy \cite{Hui:1998dc,Chung:2011hv, Chung:2011it, Hambye:2018qjv} or also via non-thermal production mechanism \cite{Falkowski:2012fb}. More recently, the mechanism of \emph{bubble filtering} (BF) \cite{PhysRevLett.125.151102,Chway:2019kft,Marfatia:2020bcs} was proposed as a way to go around the GK bound and produce ultra-heavy DM candidate with the observed abundance.

In this paper, we would like to present a new mechanism of DM 
production, occurring during strong FOPT's with ultra-relativistic walls and effective {when DM is connected via \emph{portal} coupling to the sector with FOPT.} In \cite{Vanvlasselaer:2020niz}, 
authors showed that an ultra-relativistic wall, with Lorentz 
factor $\gamma_w \gg 1$, sweeping through the plasma can excite
degrees of freedom of mass up to $M \sim \sqrt{\gamma_w} 
T_{\text{nuc}}$, possibly producing out-of-equilibrium 
particles, mechanism that we call {\emph{Bubble Expansion}} 
(BE) production. In this paper, we would like to show that 
those produced particles can be stable and thus constitute 
viable DM candidates. In addition to the possibility of evading
the GK bound and thus possibly providing ultra-massive DM 
candidate, the relic density of these particles is set by the 
hierarchy between the mass of the DM and the scale of the 
transition and thus evades the exponential sensitivity 
typically {showing up in the relic abundance controlled by FOPT's}. 

In this context, a simple model for the DM sector perhaps is a 
real singlet scalar field stabilized by a $Z_2$ symmetry 
{coupled via the \emph{portal} coupling  to the scalar field (Higgs) undergoing FOPT}. In this minimal 
setting {if the Higgs is SM field}\cite{Silveira:1985rk, McDonald:1993ex,Burgess:2000yq, 
Cline:2013gha, Abe:2014gua,Arcadi:2019lka}, FO mechanism is under strong constraints and the direct detection experiment have excluded most of the parameter region below the TeV range.

A similar production mechanism, the Bubble Collision mechanism, takes advantage of the large excursion of Higgs vacuum expectation value (VEV) during the collision of relativistic bubbles. It was first hinted in \cite{Watkins:1991zt}, predicting a production of particles as massive as $ M \sim \gamma T$. This was shown to be too optimistic in \cite{Falkowski:2012fb} where only the vector and fermion DM candidate were considered as promising DM candidate, 
{however for the \emph{scalar} DM we find that the mechanism of production of DM via bubble collision of \cite{Falkowski:2012fb} is completely subdominant compared to BE, presented in this paper.}

{We will show that our production mechanism can proceed even with very massive DM candidate}, thus possibly evading the direct detection experiment bounds, even if the coupling to SM is strong. However an irreducible prediction of the mechanism, which takes advantage of a strong FOPT, is the large imprint left in the Stochastic Gravitational Waves Background (SGWB).
Such SGWB signal could be detected in forthcoming GW detectors such as LISA, advanced LIGO, BBO, DECIGO, etc, offering an alternative way to study DM production. 
 
This paper is organised as follows: in Section \ref{sec:prod} we present the production mechanism and the amount of relics produced after the passage of the wall. In Section \ref{sec:Gen}, we present first the maximal amount of DM abundance that can be produced via BE mechanism, and then discuss three ways of accommodating the parameter space to the observed DM abundance; \ref{sec:co_ann}, we discuss how  annihilation can modify the early relics abundance, in Section \ref{sec:baby_zillas} we discuss how some amount of supercooling modifies the relative FO and BE abundances and, finally, in Section \ref{sec:loinflation}, we discuss the case of very massive DM candidate in the absence of FO relics. In Section \ref{sec:EWPT}, we specialize to the Electroweak Phase transition (EWPT) and discuss the allowed range of parameter providing the observed relic abundance. In Section \ref{sec:signature}, we expose the unavoidable gravitational signature expected by such mechanism. Finally, in Section \ref{sec:conc} we conclude. 

\section{DM production in the Bubble Expansion}
\label{sec:prod}
Let us introduce the Lagrangian for the minimal model which suffices for the illustration of the advertised effect
\bea
\mathcal{L}_{h} =\d_\mu h \d^\mu h^\dagger - V(h),
\label{eq:Lag}
\eea
where $h$ is a complex scalar field obtaining a non-vanishing VEV via the phase transition and $V(h)$ is its potential. We will not specify the form of $V(h)$ in this paper, but will assume that it leads to the first order phase transition in the early universe.
This field $h$ can be the physical Higgs, and thus the phase transition(PT) is electroweak (EWPT), or a new Dark Higgs, 
and then the transition happens only in the Dark Sector 
(DS). On the top of it, we introduce a DM candidate $\phi$, 
that for simplicity we take to be only a single scalar field
stabilized by a $Z_2$ symmetry, with Lagrangian of the form
\bea
\mathcal{L}_{\phi, h} = \frac{1}{2}(\d_\mu \phi)^2  - \frac{M^2_\phi}{2}\phi^2 - \frac{\lambda}{2} \phi^2 |h|^2.
\label{eq:phi_lag}
\eea
We have assumed that DM candidate is coupled to the symmetry breaking sector via the portal coupling which is  the simplest and most natural non-gravitational connection between the symmetry breaking sector and the DM candidate (for review on portal DM, see \cite{Arcadi:2019lka}). {We will also assume $\lambda>0$ in order to make sure the potential is bounded from below.} 
 
 We will be mostly interested in masses of the DM candidate $\phi$ much larger than the Higgs scale, $M_\phi \gg m_h$. As a consequence, the abundance of $\phi$ in the plasma at the moment of the transition is \emph{Boltzmann-suppressed} and can be ignored in
 the dynamics of the transition. We thus neglect the 
 quartic part of $\phi$ potential in the discussion 
 as well as the change of $M_\phi$ due to the 
 transition $\frac{\Delta M_\phi^2}{M_\phi^2} = 
 \frac{\lambda v^2}{M_\phi^2} \ll 1$, with $v$ the VEV of the Higgs-like field in the zero-temperature true vacuum, $v \equiv \langle h\rangle$.  {The hierarchy $M_\phi \gg m_h, v$ introduces the usual tuning of the scalar mass into the model {if $\lambda M_\phi^2/(16\pi^2)\gg m_h^2, v^2$} (similar to the SM Higgs mass hierarchy problem), but in this paper we will not try to present a model where this hierarchy can be obtained naturally. }

\subsection{Dynamics of the bubble wall after nucleation}
Let us now turn to the dynamics of the transition triggered by the Higgs-like field $h$. As already stated above, we will focus on the regime with  ultra-relativistic bubble wall expansion with $\gamma_w \equiv (1-{v^2_w})^{-1/2} \gg 1$, where $v_w$ is the wall velocity at the bubble center frame. This regime is favoured when the transition is strong enough to develop at least some amount of supercooling. {Indeed the condition for the acceleration of the wall} is fulfilled if the release of energy $\epsilon \equiv \Delta V$ (using the zero-temperature potential) is larger than the pressure $\Delta \mathcal{P}$ (computed using the zero-temperature minima) exerted on the wall by the plasma. In the relativistic limit, at the leading order (LO), the pressure is equal to\cite{Dine:1992wr,Bodeker:2009qy} 
\bea
\Delta \mathcal{P}_{\text{LO}} \to \sum_i g_i c_i \frac{\Delta m_i^2}{24}T_{\text{nuc}}^2,
\label{eq:LOpresLO}
\eea
with $g_i$ the number of degrees of freedom (d.o.f.) contained in the plasma\footnote{At this point, let us notice that if the DM candidate is decoupled from the plasma, it will not induce pressure via this mechanism}, $\Delta m^2 = m^2_{\text{broken}}- m_{\text{symmetric}}^2$ and $c_i = 1(1/2)$ for bosons (fermions) and $T_{\rm nuc}$, the nucleation temperature, is the temperature when there is roughly one bubble per Hubble volume. Eq.\eqref{eq:LOpresLO} can be considered as an upper bound on the pressure\cite{Mancha:2020fzw} and the bubble becomes relativistic if\cite{Espinosa:2010hh}
\bea
\epsilon >\Delta \mathcal{P}_{\text{LO}} \quad ({\text{Relativistic wall condition}}).
\label{eq:rel_cond}
\eea
As pressure scales like $ \mathcal{P}_{\text{LO}} \propto v^2 T^2$ and release of energy like $\epsilon \propto v^4$, supercooled transitions, like in nearly conformal dynamics\cite{Konstandin:2011dr,Bruggisser:2018mrt,Azatov:2020nbe}, drive the wall to ultra-relativistic regimes. Note that if no other contribution is present, the bubbles satisfying Eq.\eqref{eq:rel_cond} become \emph{runaway} (permanently accelerating). If some gauge field acquires a mass during the phase transition, it is known that the Next-To-Leading order (NLO) correction to the pressure\cite{Bodeker:2017cim}, due to the emission of ultra-soft vector bosons, scales like $\gamma_w$
\bea
\Delta\mathcal{P}_{\text{NLO}} &\simeq & g_i g_{\rm{gauge}}^3\gamma_w T_{\text{nuc}}^3 \frac{v}{16\pi^2},
\label{eq:LOpresNLO}
\eea
where $g_{\rm{gauge}}$ is the  gauge coupling and $g_i$ counts the number of degrees of freedom. This pressure will %thus finally 
stop the acceleration of the wall and wield a \emph{terminal} velocity with final boost factor $\gamma_{w,\text{MAX}}$. 
 Before proceeding further let us estimate the maximal velocity (or $\gamma_{w,\text{MAX}}$ factor) the bubble wall will reach before the bubble collisions.  As we have seen from Eqs. \eqref{eq:LOpresLO} and \eqref{eq:LOpresNLO}, the discussion changes depending on the presence of phase-dependent vectors.
\begin{enumerate}
\item {\bf Runaway regime:} When the PT does not involve phase-dependent vectors, there is no NLO pressure and the wall keeps accelerating until collision. The $\gamma_w$ at collision is
\cite{Azatov:2019png, Vanvlasselaer:2020niz} 
{
\bea
&&\gamma_{w,\rm MAX} \simeq \frac{2 R_*}{3 R_0}\l(1-\frac{\mathcal{P}_{\rm LO}}{\epsilon}\r), \quad R_0 \sim 1/T_{\text{nuc}}, \quad R_*\approx \frac{(8 \pi)^{1/3}v_w}{\beta(T_{\rm nuc})},\nonumber\\
&&\beta(T)=H T\frac{d}{dT}\l(\frac{S_3}{T}\r)\sim H\sim \frac{v^2}{M_{\text{pl}}}~~\Rightarrow \gamma_{w,\rm MAX} \sim  \frac{M_{\text{pl}}T_{\text{nuc}} }{v^2},
\label{eq:MAXboost}
\eea}
where $R_\star$ is an estimate for the bubble size at collision and $R_0$ is the bubble size at nucleation, $\beta$ is the inverse duration parameter of the transition and $M_{\text{pl}}\approx 2.4\times 10^{18}\,$GeV the reduced Planck mass.

\item {\bf Terminal velocity regime:} When the PT gives a mass to vectors, the pressure becomes dominated by the emission of ultra-soft bosons and quickly wield a terminal velocity of the form
\bea
\label{eq:gamma_max}
&&\Delta \mathcal{P}_{\text{NLO}} \simeq  g_i g_{\rm gauge}{\gamma_w T_{\text{nuc}}^3 \frac{v}{16\pi^2},} \qquad \epsilon \sim v^4 
\\ \nonumber
&&\epsilon = \Delta\mathcal{P}_{\text{NLO}} \qquad \Rightarrow \gamma_{w,\text{MAX}} \approx \text{Min}\l[\frac{M_{\text{pl}}T_{\text{nuc}} }{v^2}, \frac{16\pi^2}{g_i g_{\rm gauge}^3} \bigg(\frac{v}{T_{\text{nuc}}}\bigg)^3\r],
\eea
{where in the last step we have to take the minimal of the two values, since the bubble collision can happen before the terminal velocity regime is reached.}

\end{enumerate}

The last source of pressure is provided by the production of heavy particles   \cite{Vanvlasselaer:2020niz} including DM itself
\bea
\Delta \mathcal{P}_{\phi} \propto v^2 T^2_{\text{nuc}} \Theta(\gamma_w T_{\text{nuc}} - M_{\rm Heavy }^2 L_w).
\label{eq:Pmix}
\eea 
{Here $M_{\rm Heavy}$ is the typical mass of the heavy particles.}
This additional contribution can 
%in return
stop as well the bubbles from being in the runaway regime (see for examples \cite{Vanvlasselaer:2020niz}).

At last before we will proceed to the calculation of DM production, let us define a reheating temperature after the completion of the phase transition, which is approximately equal to
\bea
T_{\text{reh}} \approx (1+\alpha(T_{\text{nuc}}))^{1/4}T_{\text{nuc}}, \qquad \alpha(T) \equiv \frac{\epsilon}{\rho_{\text{rad}}(T)},
\label{eq:alpha}
\eea
where $\epsilon$ is the latent heat released during the FOPT. Generically we expect 
$T_{\text{reh}}  \sim T_{\text{cr}} \sim v$, with $T_{\text{cr}}$ the critical temperature when the two minima are degenerate. Note that in the regime of large supercooling $\alpha \gg 1$ there will be a hierarchy between the nucleation and reheating temperatures $T_{\rm reh} \gg T_{\rm nuc}$.

\subsection{Production of DM via the bubble wall}
After those preliminaries, we can now go to the production mechanism itself. In the wall frame, $h$ particles hit the wall with typical energy and momentum $E^{h} \sim p^{h}_z \sim \gamma_w T_{\text{nuc}}$. The VEV of the $h$ $\langle h \rangle = v(z)$, varying along the wall, induces a new trilinear coupling of the form 
$
\lambda v(z) h \phi^2
$ that did not existed on the symmetric side of the wall. It was shown in \cite{Vanvlasselaer:2020niz} that, in such a situation, the transition from light to heavy states $h \to \phi\phi$ has a probability of the form (see also Appendix \ref{app:prod} for the details of the computation)
\bea
P(h \to \phi^2) \approx \bigg(\frac{\lambda v}{M_\phi}\bigg)^2\frac{1}{24\pi^2} \Theta ( 1- \Delta p_z L_w) \simeq \bigg(\frac{\lambda v}{M_\phi}\bigg)^2\frac{1}{24\pi^2} \Theta \l( p_z-\frac{M_\phi^2}{v} \r).
\label{eq:prob}
\eea
 $L_w$ is the width of the wall which is approximately  $L_w
 \sim 1/v$ and $\Delta p_z \equiv p^h_z - p^\phi_{z,1}-p^\phi_{z,2} \approx \frac{M_\phi^2}{2p^h_z}$ is a difference of momenta 
 between final and initial state particles in the direction 
 orthogonal to the wall. 
% In the second equality we have used that 
%in the wall frame  the exchange of momentum between the incoming $h$ and the outgoing $\phi$'s particles is  $\Delta p_z \equiv p^h_z - p^\phi_{z,1}-p^\phi_{z,2} \approx \frac{M_\phi^2}{2p^h_z}$.
{Immediately after the production, the typical energy of each $\phi$ in the bubble center frame is 
\beq
\bar{E}_\f\sim \frac{M_\f^2}{T_{\text{nuc}}},
\eeq
if $p_z^h\gg M_\f$, see Appendix \ref{app:prod}. This is much larger than either $m_h$ or $M_\f$.}

As a consequence, inside of the bubble, a non-vanishing density of non-thermal $\phi$ accumulates. 
Thus, this ``\emph{Bubble Expansion} (BE)'' produced density of $\phi$, in the wall rest frame, takes the following form 
\bea
n_\phi^{\text{BE}} &\approx & \frac{2}{\gamma_w v_w}  \int \frac{d^3p}{(2\pi)^3} P(h \to \phi^2) \times f_h (p, T_{\text{nuc}}) 
\nonumber \\ 
&\approx &
 \frac{2\lambda^2 v^2}{24\pi^2 M_{\phi}^2 \gamma_w v_w}  \int \frac{d^3p}{(2\pi)^3}  \times f_h (p,T_\text{nuc})\Theta ( p_z- M_\phi^2/v),
\label{eq:density_1}
\eea 
 $v_w = \sqrt{1-1/\gamma_w^2}$ is the velocity of the wall, and $f_h(p)$ is the equilibrium thermal distribution of $h$ outside of the bubble. 
This writes $f_h (p) = \frac{1}{e^{\frac{{\gamma_w}(E_h - v_wp^{h}_z)}{T_{\text{nuc}}}}-1}$, as the Higgs-like field should  be at equilibrium with SM.

Using Boltzmann distribution as a simplifying assumption, $f_h (p) \approx e^{-\frac{{\gamma_w}(E_h - v_wp^h_z)}{T_\text{nuc}}}$ and $E_h = \sqrt{p_z^2 + \vec{p}^2_\perp}$, we can perform the integral in Eq. \eqref{eq:density_1}, obtaining
\bea
n_\phi^{\text{BE}} =   \frac{\lambda^2 }{48\pi^4 \gamma_w^3 v_w}\times  \frac{v^2T_\text{nuc}^2}{M_\phi^2}\bigg( \frac{M_\phi^2/v}{1-v_w}+ \frac{T_\text{nuc}(2-v_w)}{\gamma_w (v_w-1)^2}\bigg) \times e^{- \gamma_w \frac{M_\phi^2}{v} \frac{1-v_w}{T_\text{nuc}}}.
\label{eq:density_2}
\eea
With $\gamma_w (1-v_w) = \gamma_w - \sqrt{\gamma_w^2 - 1} \to \frac{1}{2\gamma_w}$, the density in the plasma frame, in the limit of fast walls, becomes
\bea
n_\phi^{\text{BE}} &=&  \frac{T_\text{nuc}^3}{12\pi^2} \frac{\lambda^2 v^2}{\pi^2 M_\phi^2}  e^{-  \frac{M_\phi^2}{2vT_\text{nuc} \gamma_w }}  + \mathcal{O}(1/\gamma_w) 
\label{eq:density_f}
\eea
 The factor $e^{-  \frac{M_\phi^2}{2vT_\text{nuc} \gamma_w }}$ is a consequence of $\Theta ( p_z- M_\phi^2/v)$ in the the Equation \eqref{eq:density_1}. We can see that in the limit 
 \bea
 \gamma_w > \frac{M_\phi^2}{2vT_{\text{nuc}}}
 \label{eq:range}
 \eea
 the exponential goes to one and the density becomes independent of the velocity of the wall $v_w$. The step function $\Theta(1-\Delta p_z L_w)\simeq\Theta ( p_z- M_\phi^2/v)$ is an approximation of the transition function which depends on the exact shape of the wall. We report it for different wall ansatzs in Appendix \ref{app:prod}. It is important to note that in the regime $\Delta p_z L_w \lesssim 1$ the step function presents a good approximation and the results are independent of the wall shape as expected from the Heisenberg uncertainty principle. However, if the inequality Eq.\eqref{eq:range} is not satisfied and we are in the regime 
\bea
\frac{M_\phi}{T_{\text{nuc}}} <\gamma_w < \frac{M_\phi^2}{2vT_{\text{nuc}}} \qquad \text{(Wall suppressed production)}
\label{eq:wallsuppressed}
\eea
then the wall shape effects start to become important. We discuss this wall suppression for the $\tanh$ and gaussian walls in the Appendix \ref{app:prod}. We find that generically the deviations from the naive  step function are exponentially suppressed, so that  expression in Eq.\eqref{eq:density_2} can be used as an estimate in the  transition regime Eq.\eqref{eq:wallsuppressed}. At last, for 
\bea
\gamma_w < \frac{M_\phi}{T_{{\rm nuc}}}
\eea
the particle production gets additional suppression by the usual Boltzmann factor.
From now we will keep working with expression \eqref{eq:density_1}, keeping in mind possible departure from pure exponential suppression behaviour.

%Finite-wall effects become important and the $\Theta$-function of Eq.\eqref{eq:density_1} has to be replaced by a wall-shape dependent suppression factor. This effect is discussed in Appendix \ref{app:prod} for the case of the linear and tanh walls.
%From now on, as a precise description is strongly wall-shape dependent, we will keep working with expression \eqref{eq:density_1}, keeping in mind possible departure from pure exponential suppression behaviour. 

The final {number density} of heavy non-thermal DM, {in the unsuppressed region} is of the form 
\bea
n_\phi^{\text{BE}} \approx \frac{\lambda^2 v^2}{ M_\phi^2}\frac{T_\text{nuc}^3}{12\pi^4} \Theta( \gamma_w T_{\text{nuc}} -M_\phi^2/v).
\label{eq:density_fin}
\eea
% The energy necessary for the production of those heavy states originates from the release of energy stored in the potential{\bf: it is already captured by the pressure expression: do we keep this sentence ?. }

%The energy density of the produced DM is given by 
%$\rho_\phi^{\text{BE}} \approx n_\phi^{\text{BE}} \sqrt{v T_{\rm nuc} \gamma_w}$
%which should be smaller than $v^4$ to neglect the back-reaction.  
 %from the Higgs potential $\sim (100\rm \, GeV)^4$ in the regime $M_\phi> \lambda v$, $\lambda<4\pi$ and $T\sim T_{\rm nuc} \lesssim 100\,$GeV.

From the previous discussion, we see that an ultra-relativistic wall of FOPT sweeping through the plasma will produce heavy states, via portal coupling of Eq.\eqref{eq:phi_lag}. Assuming no subsequent reprocessing (thermalisation, annihilation, dilution by inflation ...) of the relic abundance, the nowadays abundance of Bubble Expansion (BE) produced DM is given by
\bea
\Omega^{\text{today}}_{\phi,\text{BE}}h^2 = \frac{M_\phi n_\phi^{\text{BE}} }{\rho_c/h^2} \frac{g_{\star S0}T_0^3}{g_{\star S}(T_{\text{reh}})T_{\text{reh}}^3} \approx 6.3\times 10^8 \frac{M_\phi n_\phi^{\text{BE}} }{\text{GeV}}\frac{1}{g_{\star S}(T_{\text{reh}})T_{\text{reh}}^3}.
\label{eq:scaled_ab}
\eea
where $T_0$ is the temperature today, $\rho_c$ is the critical energy density and $g_{\star S0}(g_{\star S}(T_{\text{reh}}))$ is the entropy number of d.o.f. today (at the reheating temperature). 
%Let us emphasize that the temperature entering this formula is the reheating temperature, which potentially can be far away from the nucleation temperature. 
As a consequence, plugging the expression Eq.\eqref{eq:density_fin}, the final relic abundance today writes 
\bea
\Omega^{\text{today}}_{\phi,\text{BE}}h^2 \approx {5.4}\times 10^5  \times \bigg(\frac{1}{g_{\star S}(T_{\text{reh}})}\bigg) \bigg(\frac{\lambda^2 v}{ M_\phi}\bigg)\bigg(\frac{v}{\text{GeV}}\bigg)\bigg(\frac{T_\text{nuc}}{T_{\text{reh}}}\bigg)^3 \Theta( \gamma_w T_{\text{nuc}} -M_\phi^2/v),
\label{eq:relic_ab}
\eea
and we see that the produced relic abundance is controlled by the {quantities}
\bea
\frac{T_\text{nuc}}{T_{\text{reh}}}, \qquad \frac{v}{\text{GeV}}, \qquad \lambda^2\frac{v}{M_\phi}.
\label{eq:control_parameters}
\eea

So far we have shown that a bubble with Lorentz factor $\gamma_w$ sweeping through the plasma can produce massive states up to mass  $ M_\phi^2 \lesssim \gamma_w T_{\text{nuc}} /L_w  $, where $L_w \sim 1/v$ is the width of the wall. The maximal value of the $\gamma_w$ factor depends on the particle content of the theory (particularly the presence of the gauge fields) which influences the largest DM mass which can be produced. We can estimate this maximal mass by considering two generic cases of the bubble expansion.
\begin{enumerate}
\item {\bf Runaway regime:} 
According to this maximal boost factor in Eq.\eqref{eq:MAXboost}, the maximal mass {$M_\phi^{\text{MAX}}$} that can be produced, by the sweeping of the wall, scales like
\bea
M^{\rm MAX}_\phi\sim T_{\rm nuc}  \l(\frac{M_{\text{pl}}}v\r)^{1/2}.
\label{eq:MAXmass_2}
\eea
We will study Dark Sectors of this type in Section \ref{sec:Gen}. 

\item {\bf Terminal velocity regime:} 
Similar considerations from Eq.\eqref{eq:gamma_max} give
\bea
M^{\text{MAX}}_\phi\sim {\rm Min}\l[ T_{\rm nuc}  \l(\frac{M_{\text{pl}}}v\r)^{1/2}, 4\pi v \bigg(\frac{v}{T_{\text{nuc}}}\bigg)\r], 
\label{eq:MAXmass}
\eea

{where we assumed, as in the remaining of this paper, that $g_i g_{\rm gauge}^3 \sim \mathcal{O}(1)$.} {Above this maximal mass, the production of DM becomes exponentially suppressed according to $e^{-  \frac{M_\phi^2}{2vT_\text{nuc} \gamma_w }}$, as we have seen in Eq.\eqref{eq:density_f}.}
%Moreover, if the reheating temperature, the temperature after the completion of the transition, is of order of the scale of the transition, $v \sim T_{\text{reh}}$, then $M^{\text{MAX}}_\phi \sim 4\pi T_{\text{reh}} \frac{T_{\text{reh}}}{T_{\text{nuc}}}$. 
We will study a transition of this type in the context of EWPT in the Section \ref{sec:EWPT}. 
\end{enumerate}

The final relic abundance produced during BE has to compete with {the 
%irreducible 
relic abundance} coming from FO, which provides a final relic abundance roughly of the form
\bea
\Omega^{\text{today}}_{\phi,\text{ FO}}h^2 \approx 0.1 \bigg(\frac{0.03}{\lambda}\bigg)^2 \bigg(\frac{M_\phi}{100 \text{ GeV}}\bigg)^2.
\label{eq:relic_FO}
\eea
Notice that this component exists if the reheating temperature of the Universe after inflation  is higher than $M_\f$ and if  $\f$ couples to the thermal bath not too weakly so that $\f$ is produced from the thermal scatterings. We assume this component in most parts of this paper. However, we will remove this assumption in section~\ref{sec:loinflation}.

The ratio of the nucleation temperature $T_\text{nuc}$ over the reheating 
temperature $T_{\text{reh}}$ in Eq. {\eqref{eq:relic_ab}},  originates from the 
fact that the heavy particles are actually produced at the nucleation 
temperature, but that the release of energy reheat the plasma at 
$T_{\text{reh}}$, providing the new initial condition for the evolution of the
universe. Strong FOPT's are often accompanied by long supercooling and thermal 
inflation \cite{Lyth:1995ka, Yamamoto:1985rd}, leading to the 
hierarchy between $T_\text{nuc}$ and $T_{\text{reh}}$ and strong suppression of 
the abundance. We will see that this new suppression factor can be useful in the 
range of parameters where the final relic abundance is overproduced, as 
illustrated on Fig \ref{Fig:regions}: in the region II, where the BE abundance is dominant over FO, but both of them are too large to account for $\Omega^{\text{today}}_{\text{obs}}$ {and I, where FO is not large enough}. In this range, dilution related to thermal inflation can reduce the overproduced relic abundance to $\Omega^{\text{today}}_{\text{obs}}$.

\section{Dark Sector PT production of DM}
\label{sec:Gen}

\begin{figure}
\centering
\includegraphics[scale=0.5]{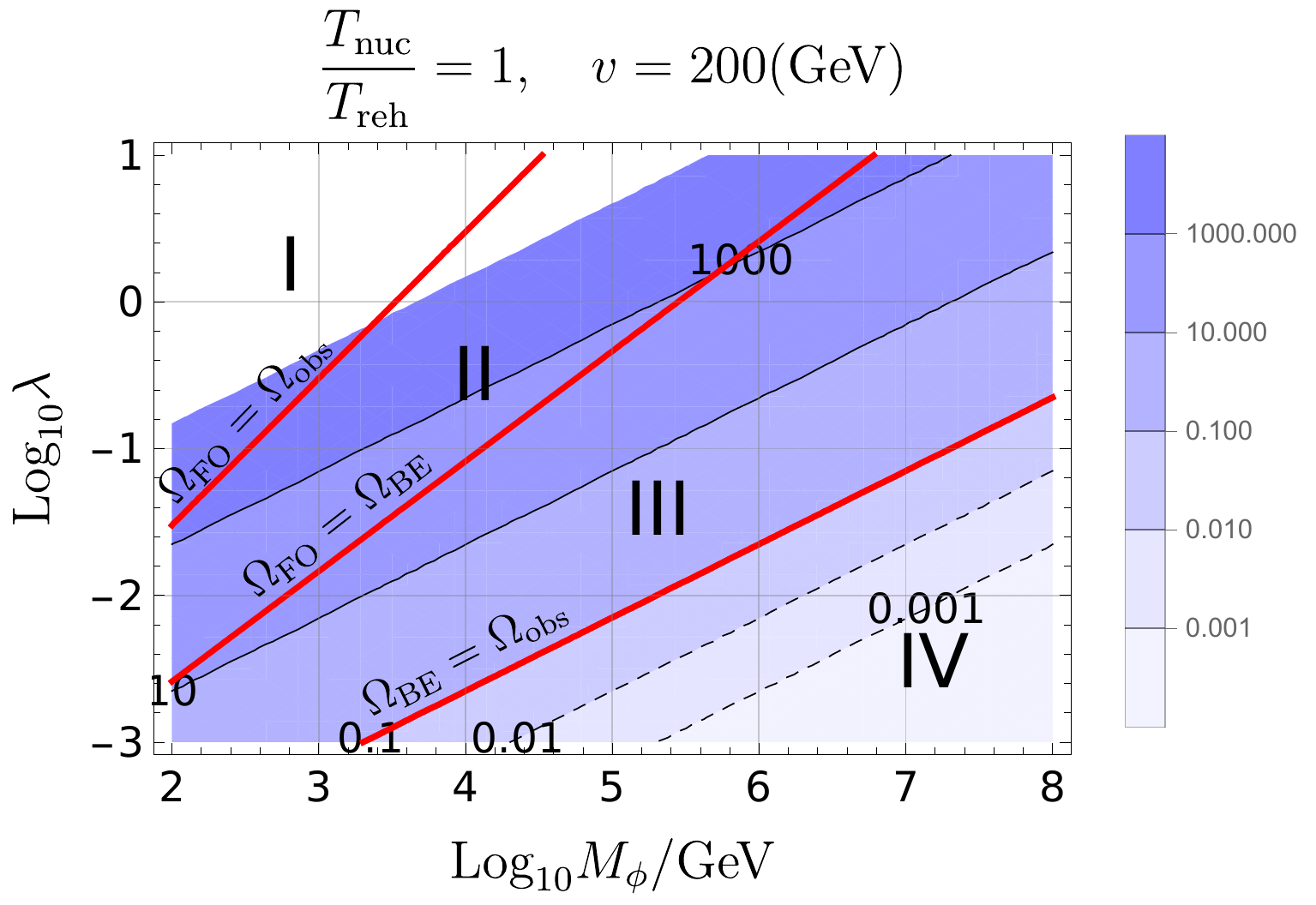}
\caption{The \emph{unprocessed} final relic abundance coming from FO and BE process with $T_{\text{nuc}} = T_{\text{reh}}$ and $v = 200$
%, 2 \times 10^4$ 
GeV. The blue shading gives the value of $\Omega_{\text{BE}}^{\text{today}}$. The red lines $\Omega^{\text{today}}_{\text{BE},\phi} = \Omega^{\text{today}}_{\text{FO},\phi}$, $ \Omega^{\text{today}}_{\text{BE},\phi} = \Omega^{\text{today}}_{\text{obs}}$ and $ \Omega^{\text{today}}_{\text{BE},\phi} = \Omega^{\text{today}}_{\text{obs}}$ define 4 regions. In I, BE abundance is dominant and FO is not enough to account for the observation. In II, FO is too large, but BE is still dominant. In III, both BE and FO are too large, but FO is dominant. Finally, in IV, FO is dominant, and BE is not enough to account for $\Omega^{\text{today}}_{\text{obs}}$.}
\label{Fig:regions}
\end{figure}

In the previous section, we have presented a new mechanism of DM production. However it is  important whether this mechanism can lead to the observed relic abundance.
In order to consider the phenomenological relevance of our mechanism we will use the toy model presented in Eq.\eqref{eq:Lag}, which can perfectly constitute a viable model of DM. We consider the field $h$ as some scalar field experiencing the phase transition
at some scale $v$. Let us look at the nowadays relic abundance presented in Eq.\eqref{eq:relic_ab}.
%$\rho_{\phi}/s$ at the instance of DM production normalized to the observed DM density {\bf hopefully this definition is correct AA ... The definition is provided in \eqref{eq:scaled_ab}}
The results are presented on the Fig. \ref{Fig:regions} for $v = 200\text{~and~} 2 \times 10^4$ GeV. 
Generically we can define four regions as follows: in region I, the abundance is under-produced 
via FO, but largely overproduced via BE. The region IV is the symmetric situation, where the 
BE is small but FO is very large. In region II(III), both FO and BE are overproduced, but BE (FO) 
production dominates over FO (BE):
\bea
I: \Omega_{BE}> \Omega_{obs},~~\Omega_{FO}<\Omega_{obs}\nn
II: \Omega_{BE}> \Omega_{FO},~~\Omega_{FO}>\Omega_{obs}\nn
III: \Omega_{BE}< \Omega_{FO},~~\Omega_{BE}>\Omega_{obs}\nn
IV: \Omega_{BE}< \Omega_{obs},~~\Omega_{FO}>\Omega_{obs}.\nn
\eea
Very naively these equations indicate that none of the regions leads to a viable phenomenology. However we have not yet taken into account few possibilities on the initial conditions as well as the evolution of { $\rho_{\f}/s$} which can make some parts of those regions viable.

To be more precise, we will study three possibilities; in the regions where DM is overproduced annihilation processes can 
reduce the DM density back to the observed relic abundance, as this can be for example the case in region I. We discuss this possibility in 
the Section \ref{sec:co_ann}.  
As we already hinted above, another process which can reduce the DM density is a brief period of inflation during the FOPT, which happens if the nucleation temperature is significantly lower than the reheating temperature. This leads to the  reduction of the overall DM density and as a result opens up some parameter space, typically inside of region I and II of Fig. \ref{Fig:regions}. We discuss this effect in the section  \ref{sec:baby_zillas}.
At last in the case that {the thermal history begins with a reheating scale below the FO temperature\footnote{This is the case that the inflaton coupling is so weak that the early produced $\f$ is diluted due to the inflaton late-time decay,
 or the inflation scale itself is low. Inflation scale can be comparable or even smaller than the weak scale in ALP inflation models~\cite{Daido:2017wwb,Takahashi:2019qmh}.   }}, $\phi$ never reaches thermal equilibrium {after the reheating} and is (almost) not produced via FO. We discuss this possibility in the Section \ref{sec:loinflation}. 
\subsection{Late time annihilation } 
\label{sec:co_ann}

 In the previous Section \ref{sec:prod}, we showed that if a relativistic bubble goes through the plasma, it can produce DM relics, possibly very over-abundant. On Fig. \ref{Fig:regions} we saw that, in region I, the FO contribution was not large enough to account for the observed DM abundance, but that on the contrary, BE production was extremely large. As a consequence we would like to track the evolution of the number of DM particles after the initial production.  We will see that, as long as the DM density produced is very large, the final density does not depend on the initial density. 
Thus the physics of this part does not change even if $\f$ is produced enormously from other dynamics e.g. inflaton/moduli decay.\footnote{An extreme scenario may be even that $\f$ is the inflaton which annihilates to reheat the Universe and becomes the DM. In this case,  we should pay careful attention to the parametric resonance.}
Due to this reason, in the following, we make a general discussion which is not specific to the BE production unless otherwise stated. 
 We assume for simplicity that the production happens instantaneously during the \emph{radiation domination} epoch at $T[t_{\rm ini}]=T_{\rm reh},$ and assume that the density just after the production is much larger than that for the observed DM abundance (which is the case of the region I of Fig. \ref{Fig:regions}).

% In the $\TEV$ mass range, 
The annihilation cross-section for the process $\f\f \to hh$ is well approximated as 
 \beq\laq{cross} \vev{\sigma_{\f\f} v_{\rm rel}} \sim \frac{g_4 \la^2 }{16\pi M_\f^2} \eeq
 when  $\f$ is non-relativistic.   
Here $v_{\rm rel}$ is the relativistic velocity, and $\vev{}$ is the average over the distribution functions of $\f$ and $h$. 
$g_4$ counts the real degrees of freedom of $h$ normalized by the number of d.o.f. of the SM Higgs doublet, $4$. For instance,  
\beq g_4=1  \AND \frac{1}{4}\eeq
for $h$ being the SM Higgs and a real singlet Dark Higgs, respectively. (In the real singlet Dark Higgs case we should take ${\cal L}_{\rm scalar}\supset -\lambda \phi^2 h^2/4.$)  
% To get the annihilation cross section, and thus the following  to the Dark Higgs, 
% we need to replace \beq \lambda^2\to \lambda^2/4 \eeq  everywhere
% because the degrees of freedoms for the final states. 
%The approximation will be very good if DM is in the thermal distribution. 
In calculating the average, we have assumed that just after the production, the DM velocity $v_\phi$ soon slows down due to the scattering with the ambient plasma, {and we further assume $h$ soon decays into the SM plasma. }
When $h$ is the SM Higgs, the assumptions are easily satisfied.  The mean-free path in the thermal environment is set by the inverse of $\G_{\rm MFP}\sim y_q^2 \frac{ \lambda^2 v^2 }{E_\f^2}T_{\rm reh}$ where $y_q$ is the quark-Higgs Yukawa coupling (This expression is valid in the broken phase. In the case of symmetric phase, the 
 scattering is with Higgs multiplet and the rate is larger.) 
 Here $T_{\rm reh}$ is comparable or larger than the mass of the quark $q.$
  $\G_{\rm MFP}$ is easily larger than the Hubble parameter unless $E_\f$ is extremely large.  
When the dominant annihilation product is a dark Higgs boson, we can still have a sub-dominant portal coupling between the DM and the SM Higgs, via which the kinetic equilibrium can be easily reached. 
 % So $\f$ soon loses it's energy, and becomes non-relativistic, 
%and gets in kinetic equilibrium with the SM plasma.  
The typical velocity of $\f$ in the kinetic equilibrium is 
 \beq
 v_\f\sim v_{\rm rel}/2 \sim \sqrt{2\frac{T_{\rm reh}}{M_\f}}.
 \label{eq:velocity}
 \eeq
 Thus a simple criterion to assess the stability of DM relics is the competition between the expansion rate of the universe, 
 $${H}[T]=\sqrt{(g_\star \pi^2 T^4/30+n_\f[t] M_\f)/(3M_{\rm pl}^2)} \approx \frac{T}{M_{\text{pl}}} T,$$ 
  and the rate of annihilation $\Gamma_{\rm ann}$. A rough \emph{stability} condition thus writes
$$ \Gamma_{\text{ann}} \sim  \vev{\sigma_{\f\f} v_{\rm rel}} n_\phi < {H} \qquad \text{(Stability condition)}.$$
 If this condition is violated the annihilation gradually takes place even if $T_{\rm reh}$ is below the FO temperature $\sim M_\phi/20,$ as discussed in  the Wino and Higgsino DM cases~\cite{Moroi:1999zb, Jeong:2011sg}.
 
\begin{figure}\begin{center}
    \includegraphics[width=75mm]{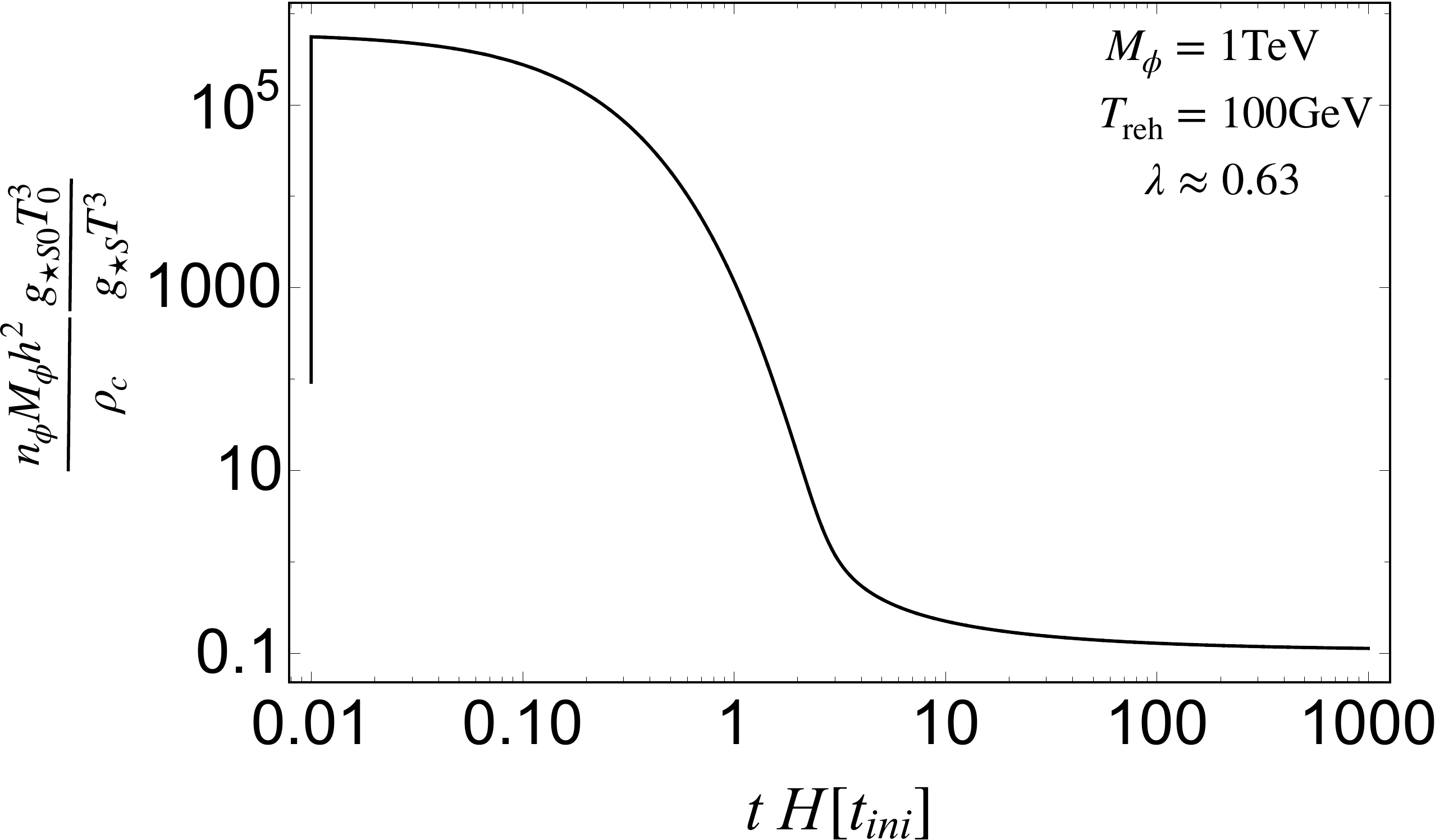}
        \includegraphics[width=75mm]{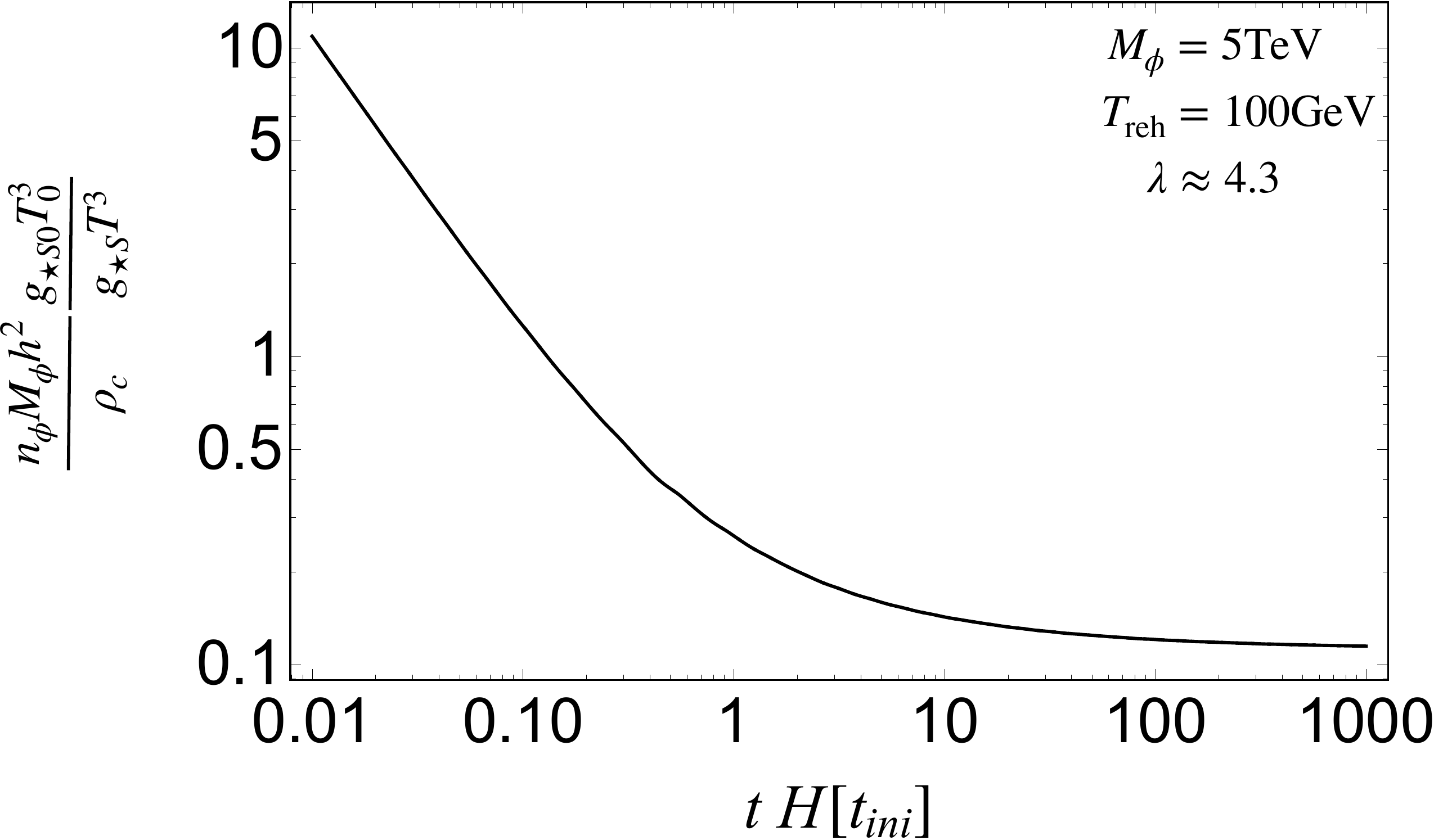}
    \includegraphics[width=75mm]{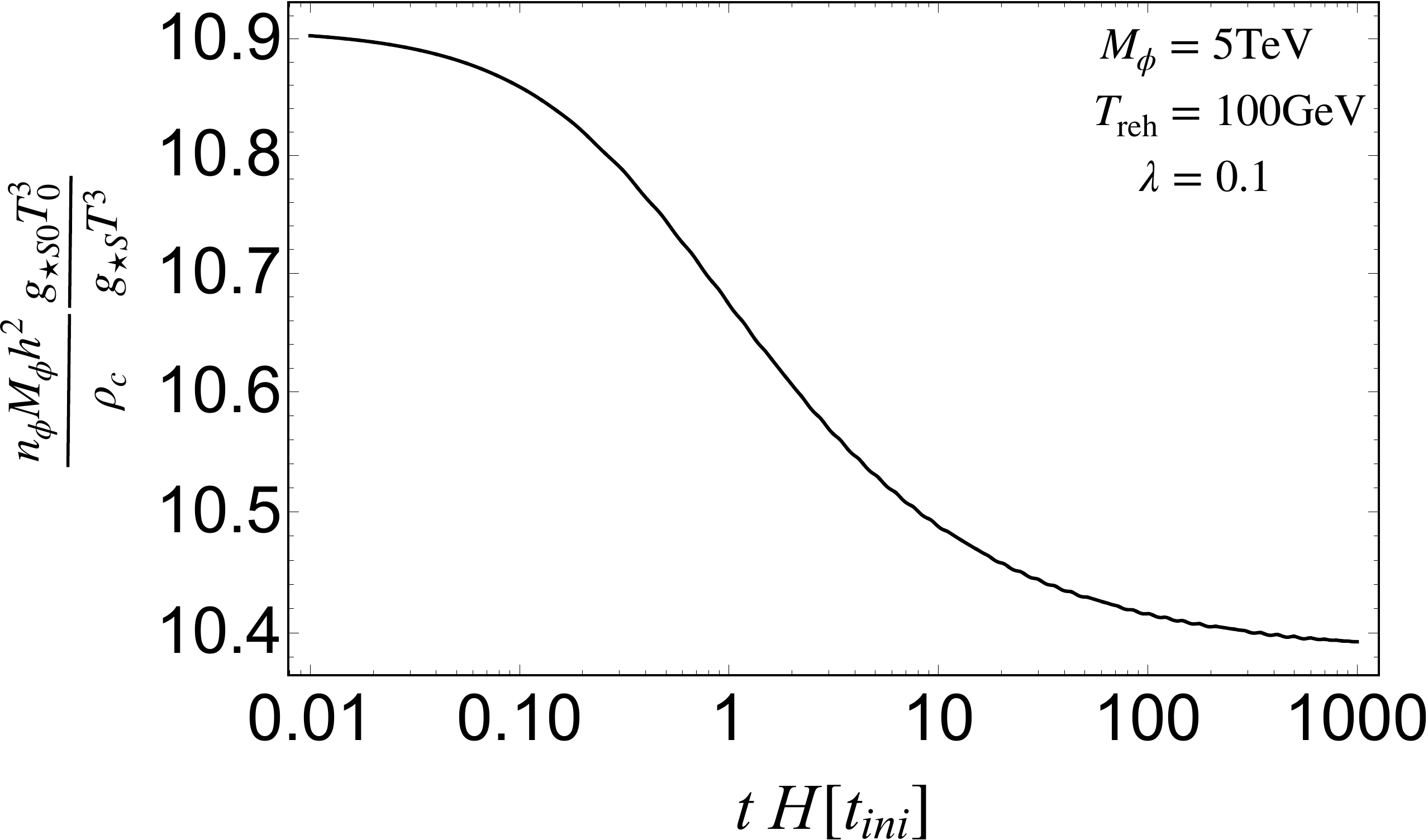}
    \end{center}
  \caption{The evolution of the energy density of the Dark Higgs portal DM, with $v=T_{\rm reh}=100\GEV,$ $M_\phi=1\TEV (5\TEV,5\TEV) $, and $\la\simeq 0.63( 4.3, 0.1)$ with large initial number density in the left top (right top, bottom) panel, which corresponds to late time FO (late time annihilation, satisfied stability condition)}
  \label{fig:Boltz} 
\end{figure}

To evaluate the final abundance after the annihilation, we can solve the integrated Boltzmann equation (by assuming kinetic equilibrium as in the case of the WIMP):
\bea\label{Boltzmanneq}
\dot{n}_\f[t]+3 \text{H} n_\f[t] = -\Gamma_{\rm ann} (n_\f[t]-n_{\rm eq}[t]^2/n_\f[t] )
\eea
$n_{\rm eq}\simeq (M_\f T/(2\pi))^{3/2}\exp{(-M_\f/T)}$ is the number density in the equilibrium,
 and the annihilation rate is given by 
\beq
\Gamma_{\rm ann}  \simeq \vev{\sigma_{\f\f} v_{\rm rel}} n_\f S_{\rm eff}
\eeq
$\vev{}$ being the thermal average and $S_{\rm eff}$ is the \emph{Sommerfeld enhancement} factor, i.e. the boost factor from the interacting long-range force.  
We assume the force potential between the $\phi$ pair distanced by $r$ as
\beq
V(r)= -\frac{\a_{\rm med}}{r}\exp{[-m_{\rm med}/r]}
\eeq
where $\a_{\rm med}$ ($m_{\rm med}$) is the messenger coupling (mass). For the Higgs-mediated force discussed in this section, 
we have $$m_{\rm med}\simeq m_h, ~~\a_{\rm med}\simeq \frac{\lambda v^2}{2\pi  M_\f^2}.$$
%with $v \approx 174\GEV.$
The analytic approximation of the enhancement factor is given by\cite{sommerfeld, Feng:2009hw} (See also Refs \cite{Hisano:2002fk,Hisano:2003ec, Cirelli:2007xd,ArkaniHamed:2008qn})
\beq
S_{\rm eff}= \frac{\pi}{\e_v}  \frac{\sinh \left(\frac{2 \pi  \epsilon_v}{\pi^2 \e_{\rm med}/6}\right)}{ \cosh \left(\frac{2 \pi  {\epsilon_v}}{{\pi ^2 {\e_{\rm med}}/6}}\right)-\cos \left(2 \pi  \sqrt{\frac{1}{{\pi ^2 {\e_{\rm med}}}/{6}}-\frac{{\epsilon_v}^2}{\left(\pi^2 {\epsilon_{\rm med}}/6\right)^2}}\right)},
\eeq
where $\e_v=v_\f/(\a_{\rm med})$ and $\e_{\rm med}=m_{\rm med}/(\a_{\rm med} M_\f).$ 
Specifically, we have $S_{\rm eff}\to \frac{\pi \alpha_{\rm med}/v_\f}{(1-e^{ -\pi \alpha_{\rm med}/v_\f})}$ with $m_h\to 0.$

To solve numerically the Boltzmann equation, we set the initial condition of $n_\f[t_{\rm ini}]\gg 0.2\EV \times s/M_\f,$ i.e. much larger than the corresponding value of the observed DM number density. 
Here $s$ is the entropy density. The Boltzmann equation can be solved to give Fig.\ref{fig:Boltz} where we plot the time evolution of the number density with $n_\f[t_{\rm ini}]\sim 40\EV \times s/M_\f$. 
Indeed, {we find that even when initially there is too large number density, with large enough coupling (and thus large annihilation rate),}
the number density decreases significantly within one Hubble time. We obtain suppressed abundance in the end (right top panel). 
On the bottom (left top) panel we can see that if the coupling is not very large this is not the case (if $M_\phi<T_{\rm reh}/20,$ $\f$ is thermalized soon and FO happens).

\begin{figure}[t!]\begin{center}
    \includegraphics[width=100mm]{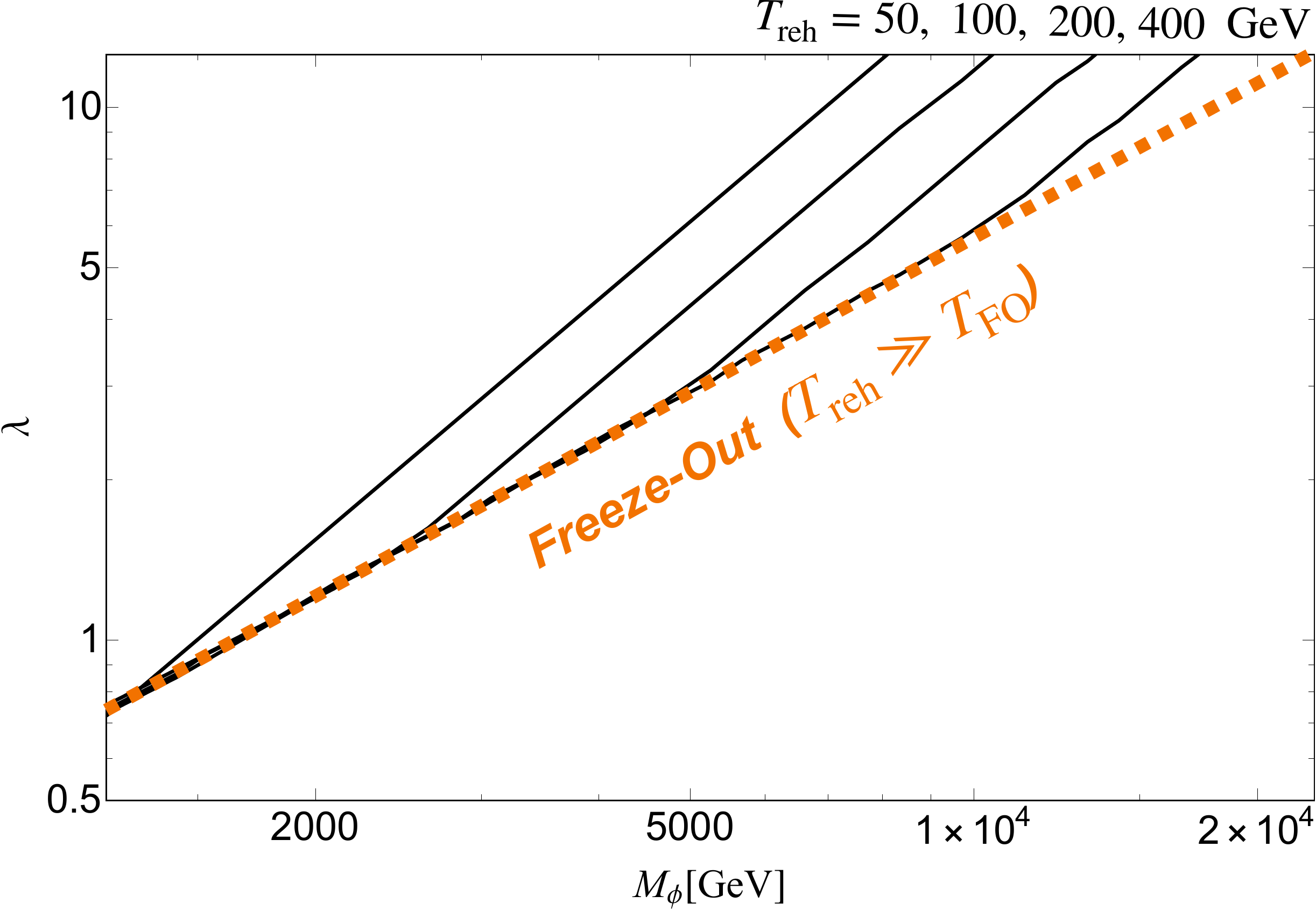}
    \end{center}
  \caption{The parameter region of the Dark Higgs portal DM with non-thermal over-production at $v=T_{\rm reh}=50\GEV,100\GEV, 200\GEV, 400\GEV$ from left to right [Black line]. $g_4=1/4.$ We neglect the mass of the dark Higgs boson. The orange dashed line indicates the FO  prediction.}
  \label{fig:regionDH} 
\end{figure}

\begin{figure}[t!]\begin{center}
    \includegraphics[width=100mm]{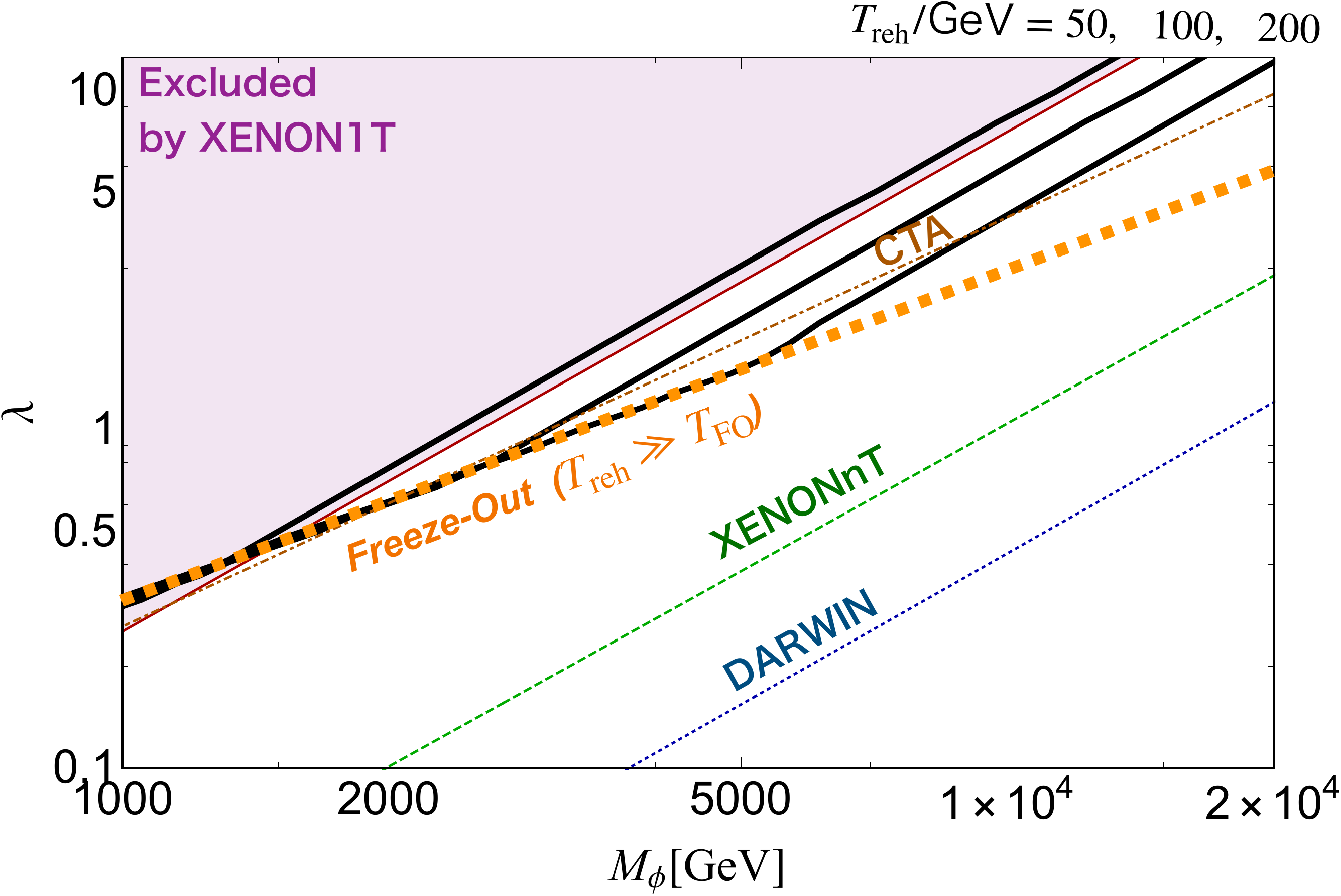}
    \end{center}
  \caption{The parameter region of the SM Higgs portal dark matter with non-thermal over production for $T_{\rm reh}=50\GEV,100\GEV, 200\GEV$ from left to right [Black line]. $v=174\,$GeV, $m_{\rm mes}=125\,$GeV, and $g_4=1.$ The orange dashed line indicates the FO  prediction. 
The purple region above the purple solid line may be excluded by XENON1T experiment \cite{Aprile:2019dbj}.
  The green dashed and blue dotted lines represent the future reaches of the XENONnT~\cite{Aprile:2020vtw} and DARWIN~\cite{Aalbers:2016jon}, respectively. The lines are adopted from \cite{Arcadi:2019lka}.
The Cerenkov Telescope Array (CTA) reach (by assuming the NFW distribution of DM) is adopted from \cite{Beniwal:2015sdl}. }
  \label{fig:regionSM} 
\end{figure}

In Fig.\ref{fig:regionDH} with $h$ being the real singlet Dark Higgs, we represent the numerical result giving $\Omega_\f h^2= \Omega_{\rm DM}h^2\approx 0.1$~\cite{Aghanim:2018eyx} by taking $v=T_{\rm reh}=50, 100, 200, 400\GEV$ from top to bottom, with the initial condition set as $n_\f[t_{\rm ini}]= 40\EV \times s/M$.  % which may correspond to different modification of the Higgs sector for PT. 
We see that at lower mass range the predictions do not depend on $T_{\rm reh}$, which represents that the FO takes over  since $T_{\rm reh}>T_{\rm FO}\sim M_\phi/20$. The FO prediction is displayed on Fig.\ref{fig:regionDH} (and \ref{fig:regionSM}) by the dotted orange line. On the larger mass range, the late time annihilation becomes important and reduces the abundance relevant to $T_{\rm reh}.$
  
In fact, we can explain the final number density, $n_\f$,  in this region from the condition 
\beq \vev{\sigma_{\f\f}v_{\rm rel}} n_\f S_{\rm eff}[T_{\rm reh}]= C H[T_{\rm reh}].\eeq 
This condition is similar to the freeze-out condition for the ordinary WIMP: the annihilation should end when the rate becomes comparable to the Hubble expansion rate. 
We obtain
 \beq
 \lambda= \lambda_{\rm ann} \sim 0.53 {\(g_4 r S_{\rm eff}\)}^{-1/2}\sqrt{C} \(\frac{g_\star(T_{\rm reh})}{103.5}\)^{-1/4} \(\frac{M_\f}{2\TEV}\)^{3/2} \(\frac{100\GEV}{T_{\rm reh}}\)^{1/2} 
 \label{eq:lamb_annihilation}
\eeq
from the condition that the $\f$ abundance composes an $r$ fraction of the observed dark matter abundance, $\Omega_\f  =r \Omega_{\rm DM}$ (and we are now focusing on $r=1$.)
Notice again that to use Eq.\eqref{eq:lamb_annihilation} we have assumed $T_{\rm FO}>T_{\rm reh}$, otherwise the DM is thermalized and then usual FO takes place after a certain redshift.  
From the numerical fit by solving the Boltzmann equations, we obtain 
$C=[0.1-1 ]$
depending on the initial condition. If the initial $n_{\f}[t_{\rm ini}]$ is larger $C$ becomes larger approaching to 1. 
In particular for our bubble wall scenario, we may have a very large $n_\f(t_{\rm ini})$ and, in this peculiar case, $C$ is almost 1.

 So far we have been agnostic regarding the coupling of the DM to the SM sectors. We just have assumed that DM couples to the scalar field $h$ to which it annihilates into.   
However, to be in \emph{kinetic equilibrium}, the DM should somehow couple to the SM plasma. This leads to the possibility of detecting DM with direct and indirect detection experiments. 
In particular, when $h$ is the SM Higgs boson, the coupling to nucleons is controlled by the coupling $\la$. 
The case where $h$ is the SM Higgs multiplet is shown in Fig.~\ref{fig:regionSM}, where the difference from Fig.~\ref{fig:regionDH} is that we fixed $v=174\GEV$, $g_4=1$ and $m_h=125\GEV$. We adopt the bound XENON1T experiment \cite{Aprile:2019dbj} from \cite{Arcadi:2019lka} (The Purple region above the purple solid line), which is extrapolated by us to multi-TeV range.  The green dashed and blue dotted lines represent the future reaches of the XENONnT~\cite{Aprile:2020vtw} and DARWIN~\cite{Aalbers:2016jon}, respectively, which  are also adapted and extrapolated from \cite{Arcadi:2019lka}. The Cerenkov Telescope Array (CTA) reach (by assuming the NFW distribution of DM) is adopted from \cite{Beniwal:2015sdl} and also extrapolated by us.
  Consequently, the predicted parameter region can be fully covered in the future direct detection and indirect detection experiments such as XENONnT, DARWIN and CTA.
{Interestingly, since the predicted black lines are parallel to the direct detection reaches in the late time annihilation region, 
$T_{\rm reh}$ corresponds to the DM-Nucleon interaction rate. If the DM is detected in the direct detection experiments, which implies the interaction rate is measured, 
we can tell the reheating temperature assuming late time annihilation. }

Here we notice that the contribution of the Sommerfeld enhancement may be as large as $S_{\rm eff}-1=\O(10\%)$  when the mass is large. 
Usually in the (SM) Higgs portal dark matter model, the Sommerfeld enhancement is negligible due to the small Higgs dark matter coupling, $\a_{\rm med}\propto (\frac{\lambda v}{M_\f})^2$ suppressed by the heavy dark matter 
mass. In the late annihilation scenario, since we need larger $\lambda$ than conventional FO and smaller $v_\f$, 
we have larger $S_{\rm eff}.$ 

As a conclusion of this section, let us, finally, come back to the BE production. We have seen on Fig. \ref{Fig:regions} that in the region of parameter with large coupling and DM mass in the TeV range, the FO is subdominant and BE is largely over-produced, this was the region I of Fig. \ref{Fig:regions}. {This is exactly the setting we studied in this section and the result displayed on Fig.~\ref{fig:regionDH} can be used for the dark sector PT.
Also Fig.\ref{fig:regionSM} can be straightforwardly extended to the EWPT, if we assume that some modification of the SM wield a strong enough EWPT. }
We will discuss this possibility further in Section \ref{sec:EWPT}.

\subsection{Dilution by supercooling}
\label{sec:baby_zillas}

In Section \ref{sec:co_ann} we saw that even if the DM is over-produced by the wall, the relic abundance can be reduced by the  reaction $\phi\phi \to hh$. For the case of $v \approx 174$ GeV, this opened up the range of values $M_\phi \in [1,10]$ TeV and $\lambda \in [0.3, 10]$, which is normally with too small abundance in usual FO. In this section, we would like to account for a second effect, which is the dilution induced by some amount of supercooling. Indeed, if some low-scale thermal inflation\cite{Lyth:1995ka, Yamamoto:1985rd} occurs due to the supercooling, a possibly large hierarchy between the reheating temperature and the nucleation temperature can occur. 

%All along this section, we assume that the DM candidate $\phi$ are already frozen-out at $T_{\text{FO}} \sim \frac{M_\phi}{20}$ before the transition, so we will be interested in the region satisfying $T_{\text{FO}} \sim \frac{M_\phi}{20} > T_{\text{reh}}$. 

During the thermal inflation\cite{Konstandin:2011dr}, the expansion factor 
scales like {$a\propto e^{H t}$ } and the temperature like {$ T_{\text{rad}}\propto e^{-
H t}$,} the FO abundance is a non-relativistic fluid scaling like $\Omega_{\text{FO}} 
\propto T^3$. As a consequence, the FO abundance receives a further $\big(\frac{T_\text{nuc}}
{T_{\text{reh}}}\big)^3$ suppression factor with respect to usual cosmology evolution.
 {Summing both  FO and BE contributions the total relic abundance will be approximately given by (we are assuming $M_\phi \gtrsim 20 T_{\rm reh}$)}
\bea
\Omega^{\text{today}}_{\phi, \text{tot}}h^2 \approx \l(\frac{T_\text{nuc}}
{T_{\text{reh}}}\r)^3 \times \bigg[\underbrace{0.1\times \bigg(\frac{0.03}{\lambda}\bigg)^2 \bigg(\frac{M_\phi}{100 \text{ GeV}}\bigg)^2}_\text{FO} + \underbrace{5\times 10^3 \times \lambda^2\frac{ v}{M_\phi}\bigg(\frac{v}{\text{GeV}}\bigg)}_\text{BE} \bigg].
\label{eq:total_ab_dil}
\eea
{When BE contribution and FO contribution are small, 
the thermal production may become dominant, especially with $T_{\rm reh}\gtrsim 1/20 M_\f$ (see Eq.\,\eqref{Boltzmanneq}). Assuming an instantaneous reheating after bubble collision and negligible non-thermal production of $\phi$ via bubble collision \cite{Falkowski:2012fb}, the additional contribution from thermal production takes the form 
\bea
\delta  \Omega_{\phi, \rm tot}^{\rm today} \sim M_\f\left.\frac{\vev{\s_{\f\f} v_{\rm rel}}n_{\rm eq}^2}{H g_{\star S}(T)T^3}  \right|_{ T=C' T_{\rm reh}}\times  \frac{g_{\star S0}T_0^3}{\rho_c} .
\label{eq:thermalprod}
\eea
This formula agrees well with the numerical simulation by taking $C'\sim 0.9-1$. 
Since, around the $T_{\text{FO}}$, this contribution changes exponentially with temperature via $n_{\rm eq},$ the range of $C'$ may be slightly wider, which depends on the detailed process of the bubble collision. 
  }
  
  {Let us also mention that, insisting on dominant BE production (second term of Eq.\eqref{eq:total_ab_dil} larger than first term and thermal production in Eq.\eqref{eq:thermalprod}), perturbativity $\lambda < 4\pi$, maximal mass Eq.\eqref{eq:MAXmass_2} and finally current bound on the relativistic species at BBN, impose the following constraints on the broken symmetry VEV of the (Dark-)Higgs:
  \bea
  {\text{MeV}} \lesssim v \lesssim 10^8\text{~GeV}, \quad \text{(scale range)}.
  \label{eq:scale_range}
  \eea
  }
  The upper bound is due to the quadratic dependence of the BE production on the VEV $v$ while the lower bound comes from stringent BBN bound on the number of relativistic species, which demands that our transition happens before $T \sim 1$ MeV.
% As a first example, on Fig \ref{Fig:regions3} we show that, considering one order of magnitude of cooling $\frac{T_{\text{reh}}}{T_{\text{nuc}}} = 10$ and $v = 200$ GeV leads to a large change in the relative contribution of FO and BE. On Fig.\ref{Fig:regions3} we display again the three red lines of Fig\ref{Fig:regions}, but now, we enlight in Green, some region of the parameter space allowing for $\Omega_{\text{obs}}^{\text{today}}$ to originate from BE production. We now see that, for one order

% of magnitude of cooling, the range of parameter $M_\phi \in [10^3, 10^5] $ GeV and $\lambda \in [10^{-2}, 10^{-1}]$ is, in principle, allowed. 
 \begin{figure}
\centering
\includegraphics[scale=0.45]{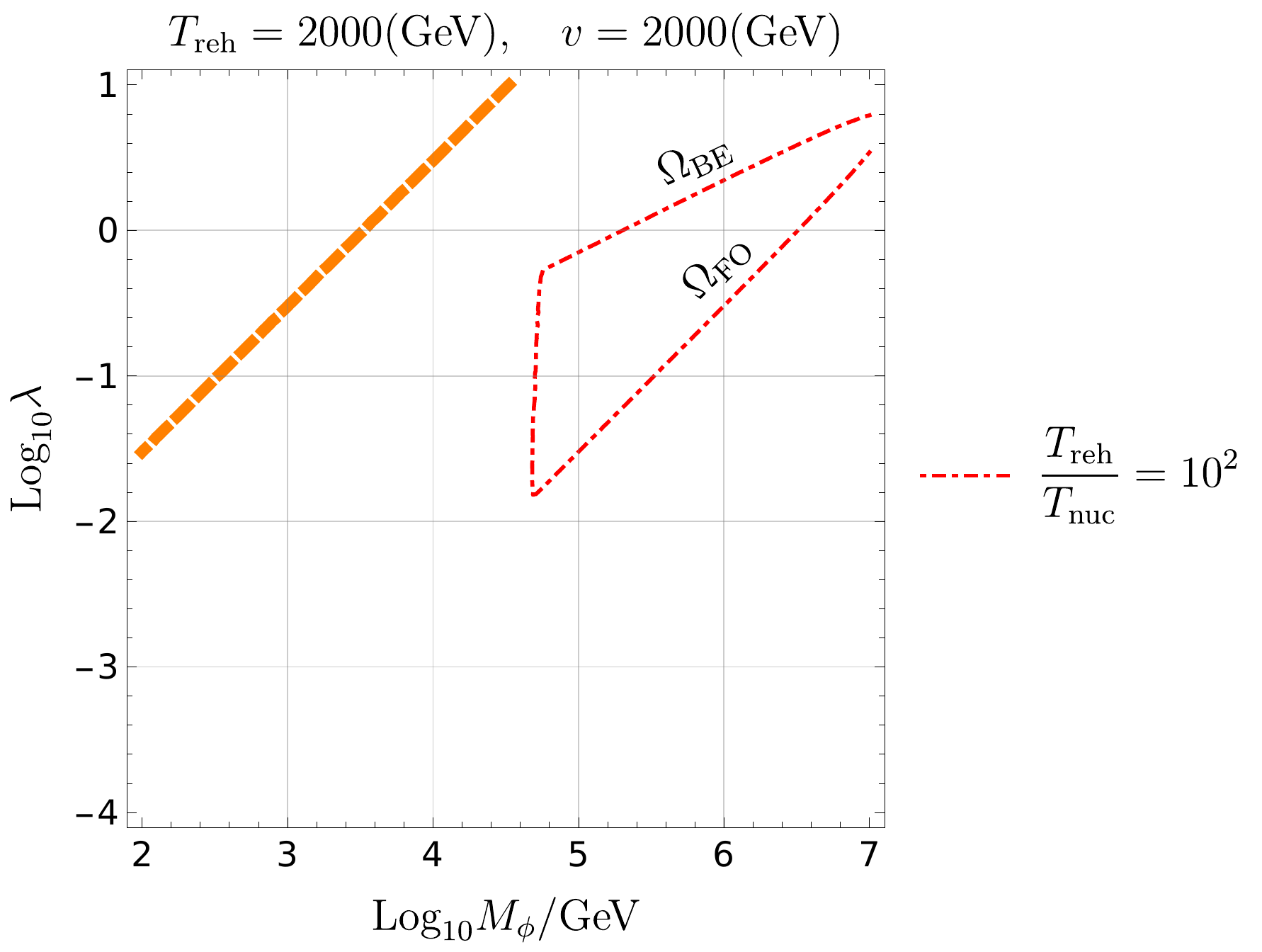}
\includegraphics[scale=0.6]{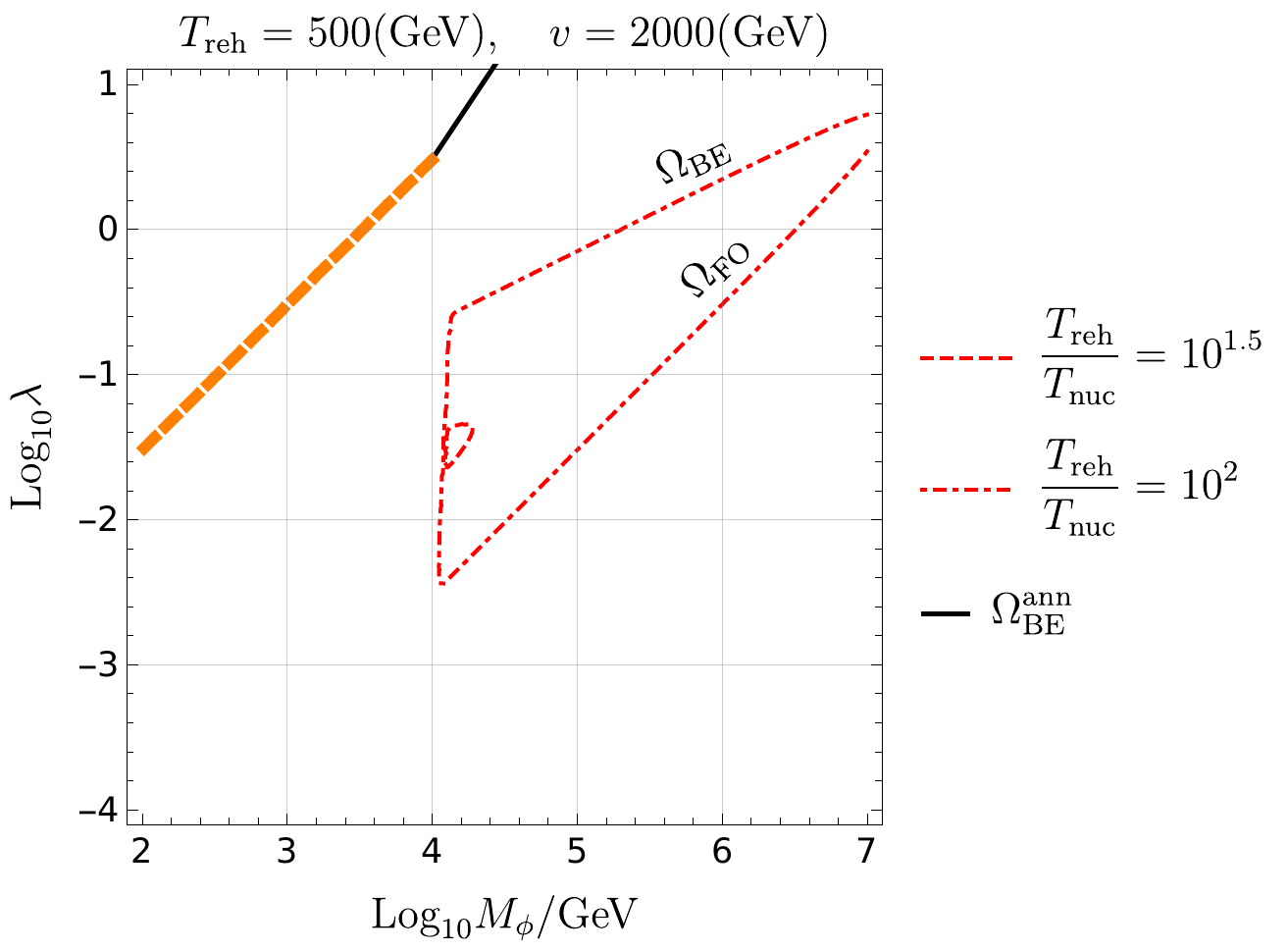}
\includegraphics[scale=0.62]{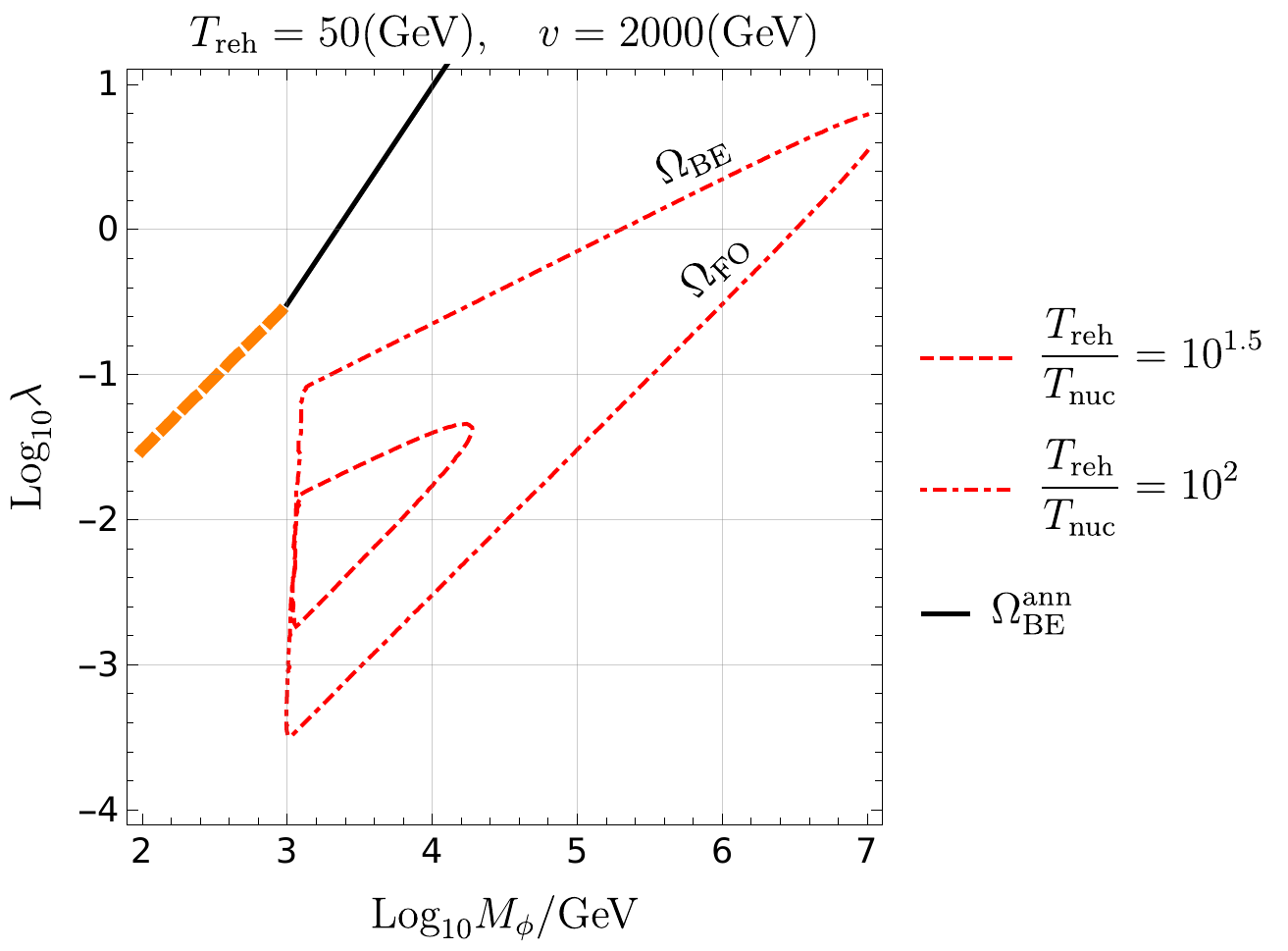}
\caption{Values of $M_\phi$ and $\lambda$ providing the observed DM relic abundance today {in the Dark Higgs portal model}, for values of supercooling $\frac{T_{\text{reh}}}{T_{\text{nuc}}} = (10, 10^{1.5},10^2)$, $v =  2000$ GeV, $g_4 = 1$. Each plot corresponds to a different value of the reheating temperature $T_{\text{reh}} = 2000, 500, 50$ GeV. The Red lines correspond to contributions from FO and BE providing the observed DM abundance and that do not undergo annihilation after the transition. The black line is the result of DM annihilation, as in Section \ref{sec:co_ann}. Roughly when $M_\phi < 20 T_{\text{reh}}$, the DM comes back to equilibrium after the transition and the final parameters compatible parameters are given by the orange dotted line. Let us also emphasize that we assumed runaway regime bubble, with the maximal DM mass given by Eq.(\ref{eq:MAXmass_2}) }
\label{Fig:regions3}
\end{figure}
On Fig. \ref{Fig:regions3}, we display the values of $M_\phi$ and $\lambda$ providing the observed amount of DM 
relics for the various values of the reheating ($T_{\rm reh}$) and nucleation ($T_{\rm nuc}$) temperatures for the fixed scale $v=2000$ GeV.
 We have also assumed that the bubble wall could reach runaway regime due to suppressed plasma pressure (no phase dependent gauge fields), 
 so that the upper bound for the DM mass in  Eq.\eqref{eq:MAXmass_2} becomes  $\sim 10^8$ GeV. These curves were obtained  by numerical solution of the Boltzmann equations but 
qualitatively we can understand the shape of the isocontours as follows:
\bit
\item Let us start with the top left plot on the Figure \ref{Fig:regions3}. The orange dashed line corresponds to the usual DM \emph{freeze-out}. As we can see, it is the case if the DM is lighter than roughly $20 T_{\rm reh}$ and, in this case, the physics of the phase transition plays no role in the final DM relic abundance.
\item For heavier masses the isocontours are given by the red dot-dashed triangles. The sides $\Omega_{\text{BE,FO}}$ of the triangles are fixed by Eq.\eqref{eq:total_ab_dil} and correspond to the cases when either $\Omega_{\text{BE}}$ or $\Omega_{\text{FO}}$ dominates the total relic abundance. Almost vertical side at $M\sim 20 T_{\rm reh}$ is given by Eq.\eqref{eq:thermalprod}
  and corresponds to the thermal production of DM during reheating after bubble collision. Inside the triangle the DM is under-produced and outside, it is over-produced.
\item  Let us move on to the other plots on the Fig. \ref{Fig:regions3}.  Multiple triangles correspond to the different values of supercooling.  
Finally the origin of the black line (continuation of the dashed orange line) can be traced  
back to the discussion in  Section \ref{sec:co_ann}. In this case the DM is produced by BE 
mechanism, however the large coupling leads to an efficient annihilation and the final relic 
abundance is set by Eq.\eqref{eq:lamb_annihilation} .

\eit

  \subsection{Super-Heavy Dark Matter candidate}
\label{sec:loinflation} 

{Another possibility to suppress the freeze-out (FO) density is to assume that the usual \emph{inflation reheating temperature} $T_R$ {is too low and inflaton does not decay into the dark matter, so that $\phi$ is not produced by reheating and thermal scattering process.\footnote{{We may also consider that $\f$ couples to the SM plasma via other couplings than that for the BE production. Then the FO component may be suppressed due to the large cross-section induced by the stronger couplings.  } }} At this point, we can completely decouple FO contribution and we are left only with the BE production, so the region of parameter space with large masses $M_\phi$ or small coupling $\lambda$ opens up.} This condition writes
\bea
T_R \ll T_{\text{FO}}\approx  \frac{M_\phi}{20} \quad \text{(No FO condition)}.
\eea

Going back to Eq.\eqref{eq:total_ab_dil} and assuming {$T_{\text{reh}} \approx v$}, we see that, in this scenario, the final relic abundance is now simply given by the BE contribution
\bea
\Omega^{\text{today}}_{\phi, \text{tot}}h^2 \approx 5\times 10^3 \times \lambda^2\bigg(\frac{ v}{M_\phi}\bigg)\bigg(\frac{v}{\text{GeV}}\bigg)\bigg(\frac{T_{\text{nuc}}}{v}\bigg)^3
\label{eq:total_ab_BBZ}
\eea
with four controlling parameters: $v, T_{\text{nuc}}, M_\phi$ and $ \lambda$. {Assuming vanishing supercooling in order to compute the maximal mass that can be produced}, DM with mass as high as 
\bea
&&M_\phi \approx  5\times 10^4 \lambda^2 \bigg(\frac{v}{\text{GeV}}\bigg)^2 \text{ GeV} 
\\ \nonumber
&&\lambda < 4\pi \quad \Rightarrow
M_\phi < M_\phi^{\text{MAX}} \approx 5\times 10^6\bigg(\frac{v}{\text{GeV}}\bigg)^2 \text{ GeV} 
\eea
can provide the observed DM abundance, $\Omega_{\text{BE}} = \Omega_{\text{obs}}$.
The second line was obtained by placing perturbativity bounds on the coupling, $\lambda < 4\pi$. {Let us emphasize that this maximal mass has nothing to do with the previously computed maximal mass in Eqs.\eqref{eq:MAXmass_2} and \eqref{eq:MAXmass}, where the production was suppressed by wall effects. In this case, the maximal mass originates from the fact that even in the unsuppressed region, the production scales as $\propto \frac{1}{M^2_{\phi}}$.} Of course, those very large masses can only be activated by the transition if it does not contain gauge boson, according to \eqref{eq:MAXmass_2}. As a consequence, this possibility most probably can not be realised in the context of EWPT, as the wall quickly reaches a terminal velocity. 

	{Fixing $v = 2\times 10^2$ GeV, and considering vanishing supercooling, the observed relic abundance is displayed, in the space $(M_\phi, \lambda)$ on Fig.\ref{Fig:regions} by the red line dubbed $\Omega_{\text{BE}} = \Omega_{\text{obs}}$.}

\section{BE production in EWPT}
\label{sec:EWPT}
So far, with the exception of Fig.\ref{fig:regionSM}, we have 
been general in our analysis and assumed that $h$ is a 
generic field undergoing a very strong FOPT. Let us now 
specialize to the case of EWPT with $v \approx 200$ GeV and 
assume that the transition is strong enough to induce a 
relativistic wall. During the SM-Higgs transition, gauge 
bosons W and Z receive a mass and thus contribute to the 
pressure at NLO order. Thus  the wall will  inevitably reach
a terminal velocity, which puts an upper bound on the 
maximal DM mass  $M_{{\phi}}^{\rm MAX}$, {above which the DM
production starts to become exponentially suppressed}. In 
Eq.~\eqref{eq:MAXmass}, we have seen that this maximal mass 
increases with the supercooling :
\bea
 M_\phi < M_{{\phi}}^{\text{MAX}} \sim \text{(TeV)}  \times \frac{T_{\text{reh}}}{T_{\text{nuc}}}.
\eea 
 As a consequence, we will study the $\phi$ relic abundance in the range 
 \bea
 \label{eq:rangemdm}
 (\text{TeV}, M_{{\phi}}^{\text{MAX}}) \sim (\text{TeV}, \frac{T_{\text{reh}}}{T_{\text{nuc}}}\times \text{TeV}) .
 \eea
We set the lower bound $ M_\phi^{\text{MIN}} \sim $ TeV, below which the usual FO takes over again after reheating if $T_{\rm reh}\sim 100\GEV$, and the sub-TeV WIMP Miracle is mostly excluded as mentioned in the introduction.
 We will  also  assume that $T_{\rm nuc} \gtrsim \Lambda_{{\rm QCD}}$, otherwise QCD effects can become important and trigger themselves phase transition (see for example \cite{Baratella:2018pxi,vonHarling:2017yew}),
  so that the longest supercooling will be roughly $\sim \frac{T_{\text{reh}}}{T_{\text{nuc}}} \lesssim 10^3$.
  These assumptions confine the DM candidate mass to be  in the range to  TeV $\lesssim M^{\text{MAX}}_{{\phi}} \lesssim 10^3 $ TeV, thus leaving us with a generous range of exploration. 
However this setting renders the scenario of Section \ref{sec:loinflation}, with very massive DM, difficult, so we will not consider it. In this Section, we will only consider the two mechanisms of Section \ref{sec:co_ann} and \ref{sec:baby_zillas}. The coupling $\lambda $ in the Eq.\eqref{eq:phi_lag} become the Higgs portal coupling and leads to the direct detection possibilities. 
\begin{figure}
\centering
\includegraphics[scale=0.5]{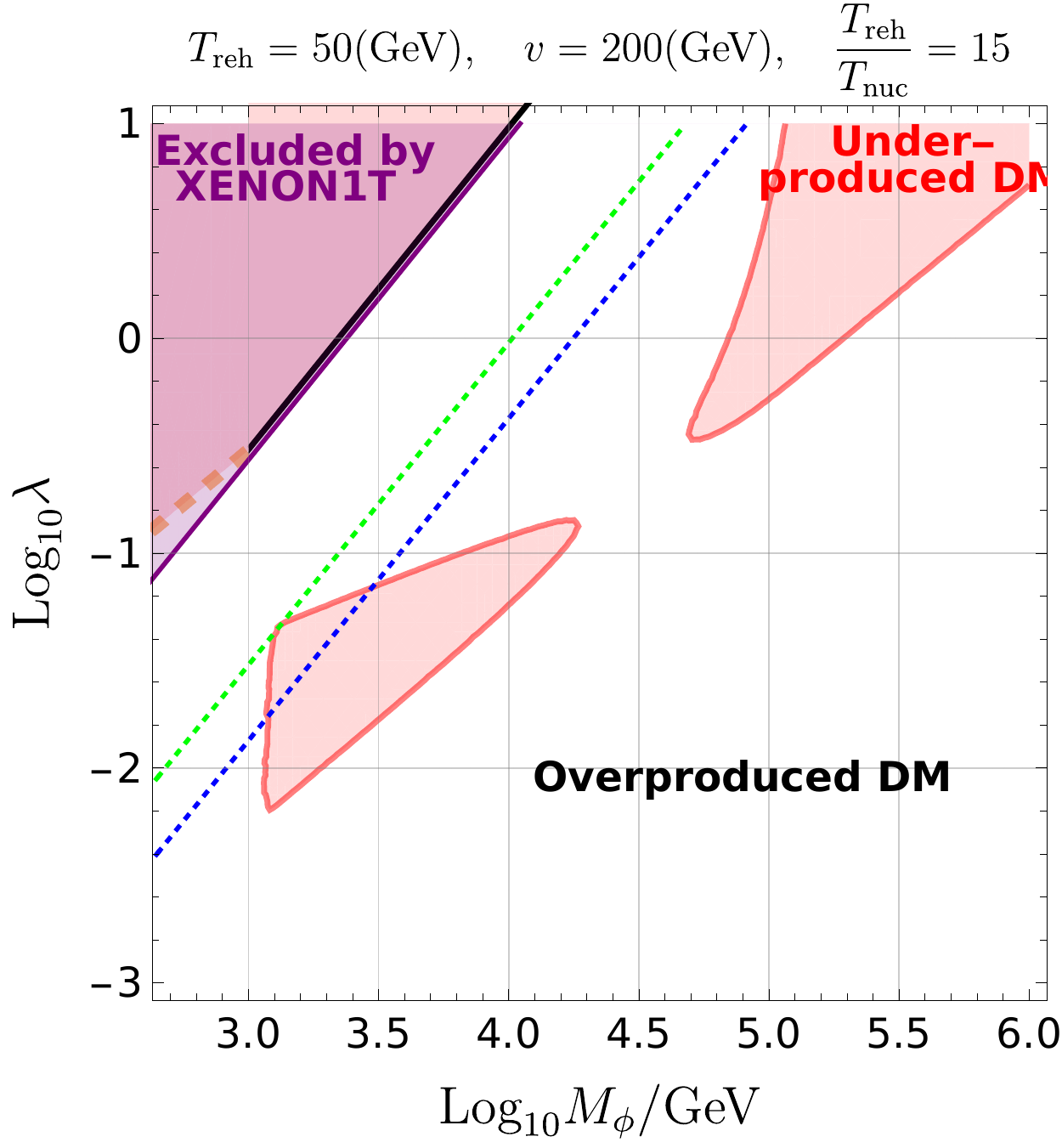}
\includegraphics[scale=0.5]{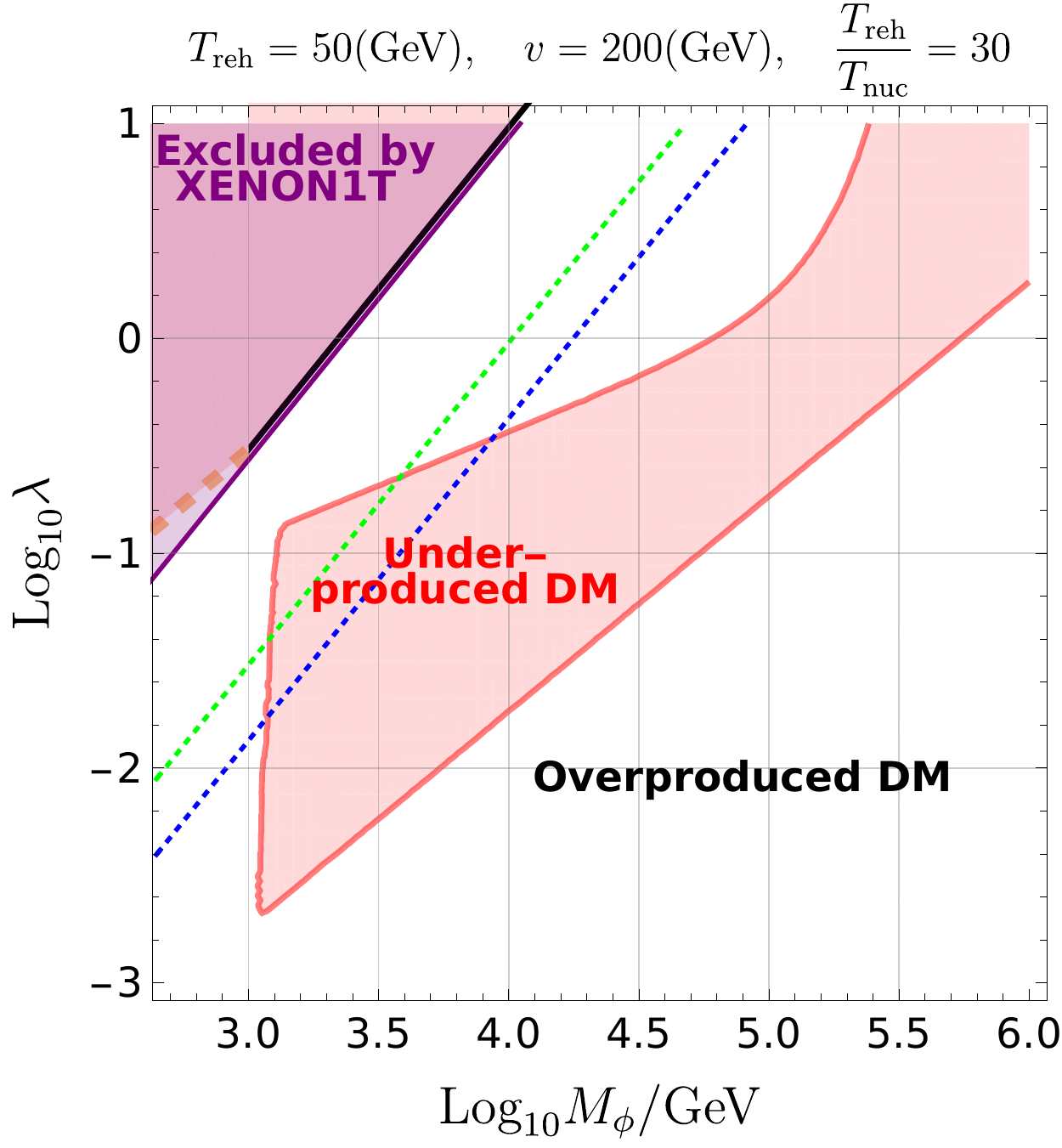}
\caption{{\bf Left}-Values of $M_\phi$ and $\lambda$ providing the observed DM abundance {in the SM Higgs portal model} for $\frac{T_{\text{reh}}}{T_{\text{nuc}}} = 15$,  $v = 200$ GeV, $T_{\text{reh}} = 50$ GeV. The orange line gives the resulting FO prediction for thermal production in the case $M_\phi > 20 T_{\text{reh}}$ and the black line is the result of DM annihilation as computed in Section \ref{sec:co_ann}. The Dotted green and blue lines are defined like in Fig.\ref{fig:regionSM}, as the future sensitivities of XENONnT and DARWIN and the violet region is already excluded by XENON1T. In the red-shaded region, DM is under-produced, outside, it is over-produced. {\bf Right}-Same plot with $\frac{T_{\text{reh}}}{T_{\text{nuc}}} = 30$. 
}
\label{Fig:regionsEWPT}
\end{figure}
Plotting the isocontours in the ($\lambda,M_{{\phi}}$) space  similarly to the Figure \ref{Fig:regions3}   we have checked the current bounds and future prospects for direct DM detection on the Fig.\ref{Fig:regionsEWPT}. We can see that parts of the parameter space where the annihilation of DM (Black line of \ref{Fig:regionsEWPT}) {plays} a role is already probed by XENON1T experiment and  parts of parameter space with BE production mechanism will be tested by the future DARWIN and XENONnT experiments, {at least partially}. 
The red-shaded region displays the regions of parameter space where the DM is under-produced, while outside of it, DM is over-produced and the observed DM abundance corresponds to the red line boundary.
It is instructive to compare these results with the  results of the Fig.\ref{Fig:regions3} where we have assumed that $\gamma_w \to \infty \Rightarrow M_{{\phi}}^{\rm MAX}\to \infty$.
On left panel of  Fig.\ref{Fig:regionsEWPT}, for $
\frac{T_{\text{reh}}}{T_{\text{nuc}}}= 15$ we can observe two islands of under-production: one 
at low mass and low coupling, which is exactly 
the same as on the Fig.\ref{Fig:regions3} and the one for the high masses and high couplings. In the later region the DM production from BE receives  an additional suppression  
 of the form $e^{-  \frac{M_\phi^2}{2vT_\text{nuc} \gamma_w }}$, according with the Eq.\eqref{eq:density_f}.
On the right panel we present a similar plot for 
 $\frac{T_{\text{reh}}}{T_{\text{nuc}}}= 30$, however in this case two islands with and without exponentially suppressed DM production are joined.
 
 Note that in our analysis we have included only the factor $e^{-  \frac{M_\phi^2}{2vT_\text{nuc} \gamma_w }}$, 
mentioned in Eq.\eqref{eq:density_2}  when we enter the regime of Eq.
\eqref{eq:wallsuppressed} and we have ignored further effects related  to the exact wall shape see discussion in the
 in Appendix \ref{app:prod} and Eq.\eqref{eq:wallsuppressed}.

To summarize  we can see that a very strong EWPT can lead to the production of a DM candidate up to {$10^2-10^3$} TeV with relatively large interaction couplings, while remaining consistent with observation.

\section{Observable signatures}
\label{sec:signature}
It is well known that an unavoidable signature of strong FOPT's, with very relativistic wall, is large a Stochastic Gravitational Waves Background (SGWB) signal, with peak frequency controlled by the scale of the transition $f_{\text{peak}}\sim 10^{-3}\frac{T_{\text{reh}}}{ \text{GeV}}$ mHz. As an example, the EWPT signal is expected to peak in the mHz range, which is the optimal range of sensitivity of the forthcoming LISA detector. 
Then the constraint Eq.\eqref{eq:scale_range} turns into a constraint on the frequency of the signal
\bea
10^{-{6}} \text{~mHz}\lesssim f_{\text{peak}} \lesssim 100 \text{~Hz}\qquad \text{(Frequency range)}
\label{eq:freq_range}
\eea
{We can also more or less constrain the model parameters for a given reheating temperature or peak frequency. In Fig.~\ref{fig:frequency}, we show  $T_{\rm reh}$ (and thus $f_{\rm peak}$ by assuming $f_{\text{peak}}=10^{-3}\frac{T_{\text{reh}}}{ \text{GeV}}$) vs the mass range. The parameter region satisfies the constraints of correct DM abundance Eq.\,\eqref{eq:total_ab_dil}$\approx 0.1$, the dominant BE production (second term of \eqref{eq:total_ab_dil} dominant, suppressed thermal production $T_{\rm reh}< 1/20 M_\phi$), Eq.~\eqref{eq:MAXmass_2}, perturbativity ($\lambda<4\pi$), and consistency conditions $T_{\rm reh}\geq T_{\rm nuc}, v \geq T_{\rm reh}.$ 
\begin{figure}
      \centering
      \includegraphics[scale=0.4]{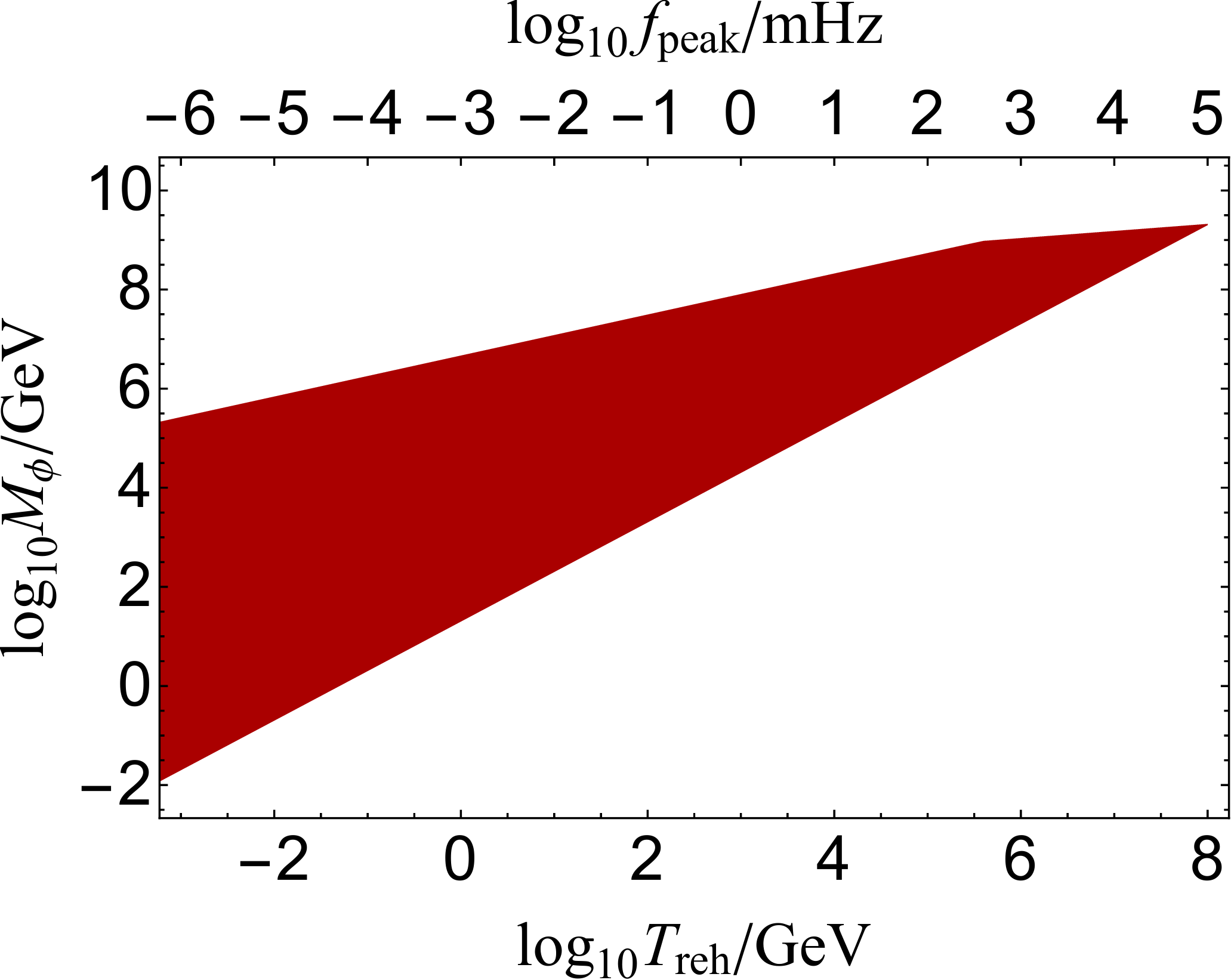}
      \caption{Reheating temperature vs the mass range of DM from BE production via a Dark PT.   Also shown is an approximate peak frequency in the upper axis. }
      \label{fig:frequency}
\end{figure} 
For the late time annihilation, we can read the relation for mass, $\lambda$, and $T_{\rm reh}$ from Figs.\,\ref{fig:regionDH} and \,\ref{fig:regionSM}.
These imply that the observation of the SGWB provides a probe of the parameter range. }

Theoretically, two different sources of GW are well understood; the \emph{bubble collision}\cite{Cutting:2018tjt}, dominating the signal in the case of runaway walls (theories with no gauge bosons), and the \emph{plasma sound wave}\cite{Caprini:2019egz}, dominating in the case of terminal velocity walls, (theories with gauge bosons). Those two contributions have peak intensity and peak frequency of the form\cite{Cutting:2018tjt,Caprini:2019egz}\footnote{As this is mostly an example, we focus on the case when shock formation is longer than Hubble time $1/H$.}
\bea
\Omega_{\text{collision}}^{\text{peak}}h^2 \sim 5\times 10^{-8} \bigg(\frac{100}{g_\star}\bigg)^{1/3}\bigg(\frac{\kappa_{\text{wall}}\alpha}{1+\alpha}\bigg)^2 (H_{\text{reh}}R_\star)^2,  
\\ \nonumber
 f_{\text{peak}} \approx 1.6 \times 10^{-4}\bigg(\frac{T}{100 \text{ GeV}}\bigg)\bigg( \frac{g_\star}{100}\bigg)^{1/6}\bigg(\frac{3.2}{2\pi H_{\text{reh}} R_\star}\bigg) \text{ Hz} 
\eea
\bea
\Omega_{\text{plasma}}^{\text{peak}}h^2 \sim 0.7\times 10^{-5} \bigg(\frac{100}{g_\star}\bigg)^{1/3}\bigg(\frac{k_{sw}\alpha}{1+\alpha}\bigg)^2 (H_{\text{reh}}R_\star),
\\ \nonumber
 f_{\text{peak}} \approx 2.6\times 10^{-5}\bigg(\frac{1}{H_{\text{reh}} R_\star}\bigg)\bigg(\frac{z_p}{10}\bigg)\bigg(\frac{T}{100 \text{ GeV}}\bigg)\bigg(\frac{g_\star}{100}\bigg)^{1/6} \text{ Hz} 
\eea
with $z_p \sim 10$, $k_{sw}$ is the efficiency factor for the production of sound waves in the plasma, $\alpha$ and $\beta$ have been defined in Eqs.\eqref{eq:alpha} and {\eqref{eq:MAXboost}} respectively, $R_\star \sim v_w/\beta \sim \mathcal{O}(10^{-2}-10^{-3})H^{-1}$ is the approximate size of the bubble at collision, and all quantities $(T,H,g_\star)$ have to be evaluated at \emph{reheating}. The specific values of the parameters $\kappa_{\text{wall}}$ and $\kappa_{sw}$ depend on the regime of the bubble expansion:
\begin{itemize}
\item {\bf Runaway wall}
A large fraction of the energy is stored in the wall of the bubble:\cite{Espinosa:2010hh}
\bea
\quad \kappa_{\text{wall}} = 1 - \frac{\alpha_{\infty}}{\alpha}, \quad \alpha_{\infty} \equiv \frac{ \mathcal{P}_{\text{LO}}+ \mathcal{P}_{\phi}}{\rho_{\text{rad}}}, \quad \kappa_{sw} \approx (1-\kappa_{\text{wall}} ) \frac{\alpha_{\infty}}{0.73+ 0.083\sqrt{\alpha_{\infty}} + \alpha_{\infty}}.
\eea
This regime produces GW via bubble collision and sound waves mechanism, with bubble collision dominating the signal.
\item {\bf Terminal velocity}
In this case, most of the energy of the transition goes to the plasma motion and we have
\bea
\quad \kappa_{\text{wall}} \to 0, \quad \kappa_{sw} \approx  \frac{\alpha}{0.73+ 0.083\sqrt{\alpha} + \alpha}.
\eea
As a consequence GW are dominated by sound wave production. 
We can see that these two scenarios are quite exclusive:
runaway behaviour is dominated by bubble 
component and  terminal velocity - by sound waves. 
This difference in principle allows 
discrimination between the two bubble expansion scenarios.
\end{itemize}

\begin{figure}
      \centering
      \includegraphics[scale=0.5]{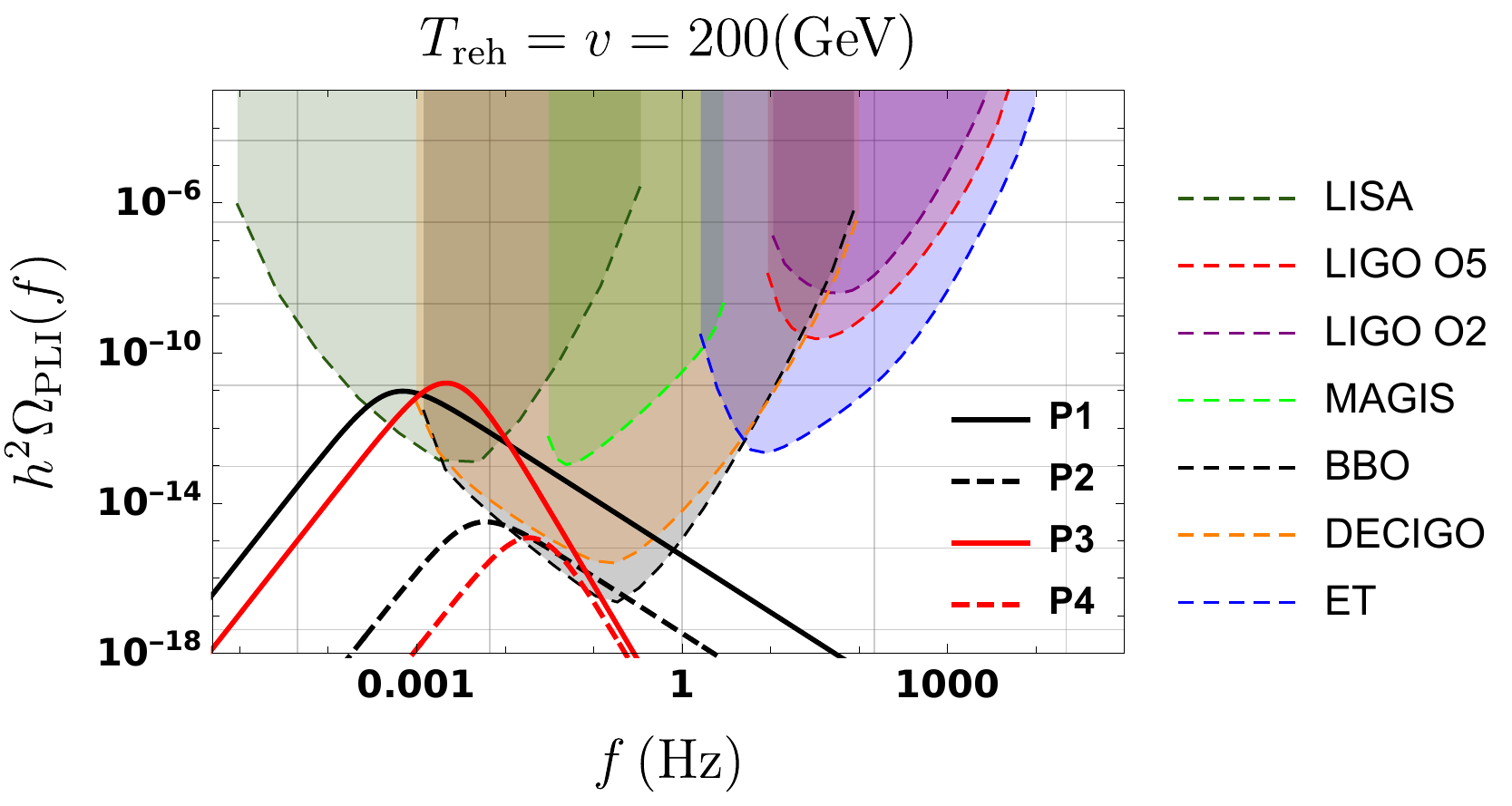}
      \includegraphics[scale=0.5]{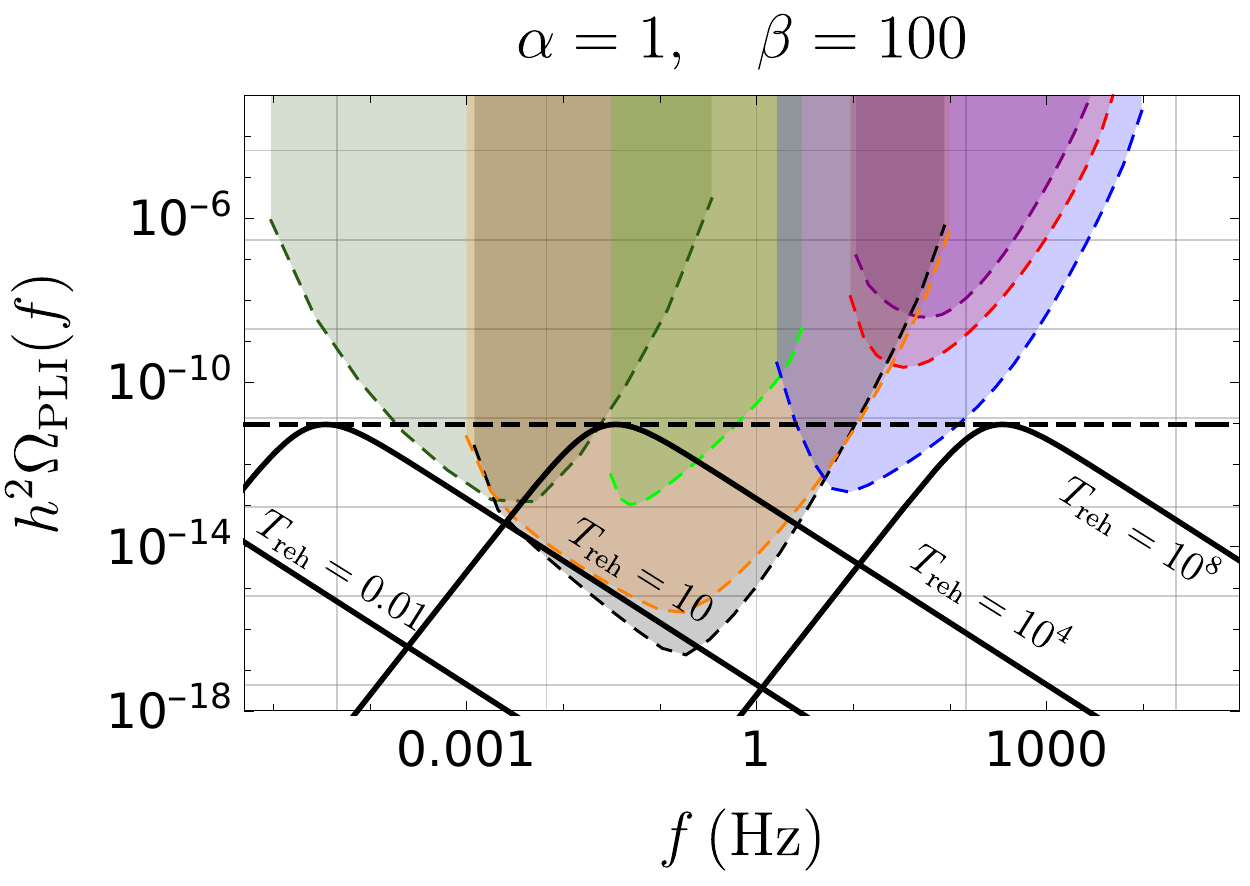}
     \caption{{\bf Left}-GW signal with $v = T_{\text{reh}} = 200$ GeV for four benchmark points in four different regimes: P1 (runaway $ \alpha = 1, \beta = 100$), P2 (runaway $ \alpha = 0.1, \beta = 1000$), P3 (terminal velocity $ \alpha = 1, \beta = 100$), P4 (terminal velocity $ \alpha = 0.1, \beta = 1000$). We also took $\alpha_{\infty} = 0.001$. The signal-to-noise ratio and the sensitivity curves can be build following the recommendations of\cite{Moore:2014lga,Aasi:2013wya,TheLIGOScientific:2014jea,Cornish:2018dyw,Graham:2017pmn,Yagi:2011yu,Yagi:2013du,Sathyaprakash:2012jk}. {\bf Right}- The runaway GW signal with fixed $\alpha=1, \beta=100$ are shown with $T_{\rm reh}=10^{-2}, 10, 10^4, 10^8 \,$GeV corresponding to the parameter range given in Fig.\,\ref{fig:frequency}.}
\label{fig:signal}
  \end{figure}

  Strong signals are obtained for: 1)large $\alpha$, which is the consequence of long supercooling and large latent heat, 2)small $\beta$, which are obtained for slow transitions and thus large bubbles at collision, and 3)relativistic walls $v_w\to 1$. Thus, the same conditions necessary for the BE production of Dark Matter will induce the strongest GW signal. In Fig.\ref{fig:signal}, we present the signal induced by four benchmark point, each representative of a specific regime:  P1 (runaway $ \alpha = 1, \beta = 100$), P2 (runaway $ \alpha = 0.1, \beta = 1000$), P3 (terminal velocity $ \alpha = 1, \beta = 100$), P4 (terminal velocity $ \alpha = 0.1, \beta = 1000$) {with $T_{\rm reh}=v=200$\,GeV. We also represent the GW signal with several $T_{\rm reh}$ in the range corresponding to Fig.\,\ref{fig:frequency} by fixing $\alpha=1 \AND \beta=100$. As we expect the scaling $\frac{\alpha_{\infty}}{\alpha} \propto \big(\frac{T_{\text{nuc}}}{v}\big)^2$, we set a suppressed $\alpha_{\infty} = 0.001$, due to quite large supercooling that we considered in most of our scenarios.}

We can see that generically BE mechanism for DM production leads to the stochastic gravitational wave signature in the frequency range Eq.\eqref{eq:freq_range}, which is well in the reach of the current and future experimental studies
\section{Conclusion}
\label{sec:conc}

	%In this paper, we presented a new mechanism to fix the relic abundance of DM. Within the very natural paradigm of \emph{portal Dark Matter}, where the DM belongs to a Dark Sector only coupled to the SM via a quartic renormalisable operator, we showed that ultra-relativistic bubble walls could produce significant additional amount of cold relics, even when the mass of the DM candidate is much larger than the scale of the transition. 
	In this paper we have presented a novel mechanism of the DM production. We have shown that  the ultra relativistic expansion of the bubbles during the first order phase transition in the early universe can  produce a significant amount of the cold relics even if the mass of the DM candidate is much larger than the scale of the phase transition. This, as a consequence, ``brings back to life'' components that, due to Boltzmann suppression, did not belong to the plasma any more. We illustrate this mechanism on a simple renormalizable model where DM is a scalar coupled via portal coupling to the  field experiencing the phase transition. When the bubble wall reaches velocities $\gamma_w >\frac{M_\phi^2}{v^2}$
the exponential suppression of the heavy particle production disappears  and BE  mechanism can become very significant in large ranges of parameter space.
Thus the produced DM density can be easily dominant. In the simple model presented in the paper 
both BE and FO contributions to the DM relic density were controlled by the
 same coupling, however this does not have to be the case  
 for more  complicated models, where additional interactions can suppress FO contribution further.

In the absence of FO produced relics, BE mechanism also provides the possibility of super-massive strongly coupled DM candidate, which is a scenario similar to the \emph{baby-zillas} of \cite{Falkowski:2012fb}.

We showed that {there are parameter regions where 
the BE production dominates over the FO production and explains the observed amount of DM in the universe. 
}
This opened up the range of Multi-TeV DM with large coupling, thus being more detectable at direct detection (like forthcoming XENONnT and DARWIN) experiments {and indirect detection (like the CTA) experiments than the usual FO mechanism.}

	Our mechanism is also characterized by an unavoidable and possibly observable imprint in the SGWB, {with peak frequency controlled by the scale of the transition}. {The shape of the spectrum can then discriminate between runaway or terminal velocity bubble wall behaviour. Let us also emphasize that if the DM belong to a totally decoupled DS, SGWB signal is the only unavoidable imprint.}
\section*{Acknowledgements}
AA in part was supported by the MIUR contract 2017L5W2PT. WY was supported by JSPS KAKENHI Grant Nos.16H06490 and 19H05810.
\appendix

 \section{Width wall effects and production of the DM}
 \label{app:prod}
 In this appendix, we explicitly work out the expression for Eq.\eqref{eq:prob}, the probability of the $1\to 2$ splitting, considering the simple model of Eq.\eqref{eq:Lag}. Usual {Poincar\'{e}} invariance would of course forbid the transition $1\to 2$. However, in the presence of the bubble wall, {Poincar\'{e}}  invariance is broken and this exotic transition can occur. We will consider the process $h\to \phi \phi$, where $h$ is the field getting a VEV, and $\phi$ is the heavy field. Assuming a bubble wall along the z direction, we define the kinematics as
\bea
p^{h}&=&(p_0,0,0,\sqrt{p_0^2-m_h^2(z)})\nn
k^\phi_1&=&(p_0(1-x),0,k_\perp, \sqrt{p_0^2(1-x)^2-k_\perp^2-M_\phi^2(z)})\nn
k^\phi_2&=&(p_0 x,0,-k_\perp,\sqrt{p_0^2 x^2-k_\perp^2- M_\phi^2(z)}).
\label{eq:kin}
\eea 
  The pressure will be now sustained by a $h \to \phi\phi$ decay in the wall. As a consequence, $M_\phi$ is (almost) independent on z, and only $m_h(z)$ is modified along the wall. 
{Here we will assume that the thermal corrections, especially the thermal mass, are neglected. This is the case for the Higgs boson with $T_{\rm nuc}\lesssim m_h$ even if the Higgs is interacting with the plasma, and is neglected for $\f$ since $\f$ is heavy. }

To estimate the probability of transition, we use the WKB method, valid as long as the incoming momentum is much larger than the length of the wall $L_w$, 
\bea
p_z \sim p_0  \gg L_w \qquad \text{(WKB condition)}.
\eea
 In this limit, the wave function takes approximately the form 
 \bea
\phi(z)\simeq \sqrt{\frac{k_{z,s}}{k_z(z)}}\exp \l(i \int_0^z k_z(z')d z'\r),
\eea
and, using the notations of \cite{Bodeker:2017cim}, the $\mathcal{M}$ matrix  writes 
  \bea 
  \nonumber
  \mathcal{M} &=& \int  \limits_{-\infty}^{\infty}dz e^{i \int\limits_0^{z} p^{h}_z(z') dz'}e^{-iq^{\phi}_z z}e^{-i k^{\phi}_z z} V(z)
  \\ 
  &\approx& \int\limits_{-\infty}^{\infty} dz e^{i p^{h}_z z}e^{-iq^{\phi}_z z}e^{-i k^{\phi}_z z} V(z) 
  \equiv \int\limits_{-\infty}^{\infty} dz e^{i \Delta p_z z} V(z),
  \label{eq:matrix_el}
  \eea
  with $ p^{h}_z(z) = \sqrt{p_0^2 - k^2_\perp - m_h^2(z)} \approx p_0$ the momentum of the incoming $h$ particle and $k^{\phi}_z, q^{\phi}_z$ the momentum of the two $\phi$ outgoing particles. In the second line, we neglected $m_h^2(z)$, as it is much smaller than $M_\phi$. We also defined $\Delta p_z \equiv p^{s}_z - q^{\phi}_z - k^{\phi}_z \approx \frac{M_\phi^2+k_\perp^2}{2x (1-x) p_0}$, the momentum exchange.
  
  To approximate the integral, we need to use some estimation for the shape of the wall. Let us approximate it using a \emph{linear}  ansatz  of the form
\bea
 \langle h\rangle =\l\{\baa{c} 0,~~z <0\\
 v\frac{z}{L_w}  ~~~0\leq z\leq L_w\\
 v ~~~z>L_w
 \eaa\r. \Rightarrow \qquad V(z) =\l\{\baa{c}  V_s \equiv 0,~~z <0\\
 \lambda v\frac{z}{L_w}  ~~~0\leq z\leq L_w\\
 V_h \equiv \lambda v  ~~~z>L_w
 \eaa\r. .
  \eea
Later we will compare with the case of more generic forms.
The integral in Eq.\eqref{eq:matrix_el} along the wall direction naturally splits into three parts
   \bea
 \mathcal{M} &=& \int\limits_{-\infty}^{0} dz e^{i \Delta p_z z} V(z)+ \int\limits_{0}^{L_w} dz e^{i \Delta p_z z} V(z)+ \int\limits_{L_w}^{\infty} dz e^{i \Delta p_z z} V(z)\nonumber
 \\ \nonumber
 &=& 0 + (1- e^{i \Delta p_z L_w} -i \Delta p_z L_w e^{i \Delta p_z L_w}) \frac{V_h}{\Delta p_z^2 L_w}  + \frac{V_h}{i\Delta p_z} (-e^{i\Delta p_z L_w} + e^{i \infty})
 \\ 
 &=& V_h\frac{1 - e^{i\Delta p_z L_w}}{L_w \Delta p_z^2}.
  \eea
  Putting together the relevant pieces, the final matrix element squared is 
  \bea
 |\mathcal{M}|^2 = \frac{V_h^2}{\Delta p_z^2} \bigg(\frac{\sin \alpha}{\alpha} \bigg)^2 =  \frac{\lambda^2 v^2}{\Delta p_z^2} \bigg(\frac{\sin \alpha}{\alpha} \bigg)^2, \qquad \alpha = \frac{L_w \Delta p_z}{2}.
 \label{eq:matrix_ele}
  \eea
With those tools in hand, we can now compute the probability of 1 to 2 splitting. The expression for the probability of transition generically takes the form
\bea
P_{h\to \phi_1\phi_2}=\prod_{i \in 1,2} \int \frac{d^3k_i}{(2\pi)^3 2 k_0^i}(2\pi)^3\delta^2 (p_\perp-\sum_{i \in 1,2} k^{i}_\perp)\delta(p_0-\sum_{i \in 1,2}  k^{i}_0)| \mathcal{M}|^2
\label{eq:Prob_trans}
\eea
and putting together Eq. \eqref{eq:Prob_trans}, \eqref{eq:matrix_ele}, using the kinematics \eqref{eq:kin} and the large velocity approximation $\Delta p_z \approx \frac{M_\phi^2+ k_\perp^2}{2x(1-x) p_0}$, we obtain
\bea
\label{eq:probtransf}
P_{h \to \phi \phi}&\simeq & \int\limits_0^{1} \frac{dx}{16p_0^2\pi^2 x(1-x)}\int d k_\perp^2 \frac{4 p_0^2 \lambda^2 v^2 x^2(1-x)^2}{(k_\perp^2+M_\phi^2)^2} \bigg(\frac{\sin \alpha}{\alpha} \bigg)^2 \Theta (p_0-2M_\phi)\nn
&\simeq &{\frac{\lambda^2}{4\pi^2} v^2}\int\limits_0^{1} dx x(1-x)\int  \frac{ d k_\perp^2}{(k_\perp^2+M_\phi^2)^2} \times \bigg(\frac{\sin \alpha}{\alpha} \bigg)^2\Theta (p_0-2M_\phi)
\\ \nonumber
&\approx &
\frac{\lambda^2}{24\pi^2}\frac{v^2}{ M_\phi^2} \times \Theta(\gamma_w {T_{\text{nuc}}} - M^2_\phi L_w)\Theta (\gamma_w T_{\text{nuc}}-2M_\phi).
\eea
where the $\Theta (\gamma_w T_{\text{nuc}}-2M_\phi)$ function appears from the trivial requirement that we need enough energy to produce the two heavy states and $\Theta(\gamma_w { T_{\rm nuc}} - M^2_\phi L_w)$ comes from the behaviour of the function $\sin \alpha /\alpha$, suppressing the transition probability for large $\alpha$. 

{ One can also estimate the typical  energy of the produced $\f$ in the bubble center frame.
\bea
\bar{E}_\f \approx \frac{1}{2}\frac{
%\left.
\int{dx x(1-x)\left[((k_1^\f)_0+(k^\f_2)_0) \gamma_w -((k_1^\f)_z+(k^\f_2)_z)v_w \gamma_w \right]}
%\right|_{z\to \infty}
}{\int{dx x(1-x)}} \sim \frac{3}{4} \frac{M_\f^2}{T_{\rm nuc}}.
\eea
Here in the last approximation we have used that $ p^h_0\sim \gamma_w (1+v_w )T_{\rm nuc} \AND v_w= \sqrt{1-\gamma^{-2}_w}$. } 
{
\subsection{Consequences of the shape of the wall}
So far, we have been assuming that the wall has a linear shape. This provided us with a suppression factor $\big(\frac{\sin \alpha}{\alpha}\big)^2$. However given a wall shape we would have different type of suppression. The wall shape depends on the Higgs potential and the specific interactions between the wall and the plasma. 
Here let us assume two types of the wall shape to explicitly calculate the suppression factor from Eq.\eqref{eq:matrix_el}.
To this end, we use tanh wall shape
\bea
V_{\text{tanh}}(z)= \frac{\lambda v}{2} \bigg[\tanh{\l(\frac{z}{L_w}\r)} +1\bigg],
\eea
and gaussian wall shape
\bea 
\label{eq:gausian}
V_{\text{gaussian}}(z)= \frac{\lambda v}{\sqrt{2 \pi} L_w } \int_{-\infty}^{z}d z'\exp{\l(-\frac{z'^2}{2L_w^2}\r)}
\eea
to perform the integral. 

In general, we can perform the integral by using partial integration of 
\bea
{\cal M}= -\int_{-\infty}^{\infty}{dz V'(z) \frac{ \exp{(i\Delta p_z z )}}{i\Delta p_z}}
\eea
where we have neglect the surface term.

For the $\tanh{(z/L_w)}$ case, $V_{\text{tanh}}'(z)= v\lambda /(2L_w \cosh^2{(z/L_w)})$. 
By noting that $z$ integral becomes the summation of residues at poles $z=  \pi i L_w/2  ,  3\pi i L_w/2 , \cdots $ for $\Delta p_z>0$ or $z= -\pi i L_w/2 ,- 3\pi i L_w/2  , \cdots $ for $\Delta p_z<0$, we obtain 
\bea
{\cal M}_{\text{tanh}}=   {\rm sign}[\Delta p_z]
\pi i \lambda v L_w \sum_{n=0}^{\infty}{ e^{-L_w |\Delta p_z| (n+ 1/2)\pi } }=\frac{ 
%{\rm sign}[\Delta p_z]{\rm sign}[\Delta p_z] 
\pi i  \lambda v L_w }{2 \sinh{\l( \frac{ L_w \Delta p_z \pi}{2}\r) }}.
\eea
One finds that this has the exactly same behavior at $\Delta p_z\lesssim 1/L_w$ but the suppression is rather exponential, $\propto e^{-L_w \Delta |p_z|}$ when $L_w |\Delta p_z|\gtrsim 1$. This implies that the linear approximation is good when $L_w |\Delta p_z|\lesssim 1$ but may not be good enough when $L_w |\Delta p_z|\gtrsim 1$.

In the case of Eq.~\eqref{eq:gausian} similarly we obtain, 
\bea
{\cal M}_{\text{gaussian}}=-\frac{\lambda v}{\sqrt{2\pi} i \Delta  p_z L_w} \int_{-\infty}^{\infty}{d z  \exp{\l(-\frac{z^2}{2L_w^2}+i \Delta p_z z\r)}}= \frac{\lambda v i }{ \Delta p_z} \exp{\l( -\frac{L_w^2\Delta p_z^2}{2} \r)}
\eea
where we have dropped the surface term. 
Again we have the same form as the linear approximation with $\Delta p_z\lesssim 1/L_w $ but the suppression factor is gaussian. 
}

\bibliographystyle{JHEP}
{\footnotesize
\bibliography{biblio}}
 \end{document}